\documentclass[10pt]{article}
\usepackage{amssymb,amsmath}
\usepackage{amsthm}
\usepackage{mathrsfs}
\usepackage{dcolumn}
\usepackage{bm}
\usepackage{fullpage}
\usepackage{color}
\usepackage[all]{xy}
\usepackage[pdftex]{graphicx,hyperref}  
\usepackage{ulem}
\usepackage{overpic}
\begin{document}
%
\theoremstyle{definition}
\newtheorem{Definition}{Definition}[section]
\newtheorem{Theorem}[Definition]{Theorem}
\newtheorem{Proposition}[Definition]{Proposition}
\newtheorem{Lemma}[Definition]{Lemma}
\newtheorem*{Proof}{Proof}
\newtheorem{Example}[Definition]{Example}
\newtheorem{Postulate}[Definition]{Postulate}
\newtheorem{Corollary}[Definition]{Corollary}
\newtheorem{Remark}[Definition]{Remark}
\newtheorem{Interpretation}[Definition]{Interpretation}
\newtheorem{Claim}{Claim}
\newcommand{\beq}{\begin{equation}}
\newcommand{\beqa}{\begin{eqnarray}}
\newcommand{\eeq}{\end{equation}}
\newcommand{\eeqa}{\end{eqnarray}}
\newcommand{\non}{\nonumber}
\newcommand{\lb}{\label}
\newcommand{\fr}[1]{(\ref{#1})}
\newcommand{\bb}{\mbox{\boldmath {$b$}}}
\newcommand{\bbe}{\mbox{\boldmath {$e$}}}
\newcommand{\bt}{\mbox{\boldmath {$t$}}}
\newcommand{\bn}{\mbox{\boldmath {$n$}}}
\newcommand{\br}{\mbox{\boldmath {$r$}}}
\newcommand{\bC}{\mbox{\boldmath {$C$}}}
\newcommand{\bp}{\mbox{\boldmath {$p$}}}
\newcommand{\bx}{\mbox{\boldmath {$x$}}}
\newcommand{\bF}{\mbox{\boldmath {$F$}}}
\newcommand{\bT}{\mbox{\boldmath {$T$}}}
\newcommand{\bQ}{\mbox{\boldmath {$Q$}}}
\newcommand{\bS}{\mbox{\boldmath {$S$}}}
\newcommand{\balpha}{\mbox{\boldmath {$\alpha$}}}
\newcommand{\bomega}{\mbox{\boldmath {$\omega$}}}
\newcommand{\ve}{{\varepsilon}}
\newcommand{\e}{\mathrm{e}}
\newcommand{\met}{\mathrm{met}}
\newcommand{\eq}{\mathrm{eq}}
\newcommand{\B}{\mathrm{B}}
\newcommand{\E}{\mathrm{E}}
\newcommand{\G}{\mathrm{G}}
\newcommand{\R}{\mathrm{R}}
\newcommand{\Z}{\mathrm{Z}}
\newcommand{\HH}{\mathrm{H}}
\newcommand{\I}{\mathrm{I}}
\newcommand{\ddiv}{\mathrm{div}}
\newcommand{\II}{\mathrm{II}}
\newcommand{\hF}{\widehat F}
\newcommand{\hL}{\widehat L}
\newcommand{\tA}{\widetilde A}
\newcommand{\tB}{\widetilde B}
\newcommand{\tC}{\widetilde C}
\newcommand{\tL}{\widetilde L}
\newcommand{\tK}{\widetilde K}
\newcommand{\tX}{\widetilde X}
\newcommand{\tY}{\widetilde Y}
\newcommand{\tU}{\widetilde U}
\newcommand{\tZ}{\widetilde Z}
\newcommand{\talpha}{\widetilde \alpha}
\newcommand{\te}{\widetilde e}
\newcommand{\tv}{\widetilde v}
\newcommand{\ts}{\widetilde s}
\newcommand{\tx}{\widetilde x}
\newcommand{\ty}{\widetilde y}
\newcommand{\ud}{\underline{\delta}}
\newcommand{\uD}{\underline{\Delta}}
\newcommand{\chN}{\check{N}}
\newcommand{\cA}{{\cal A}}
\newcommand{\cC}{{\cal C}}
\newcommand{\cD}{{\cal D}}
\newcommand{\cF}{{\cal F}}
\newcommand{\cH}{{\cal H}}
\newcommand{\cI}{{\cal I}}
\newcommand{\cK}{{\cal K}}
\newcommand{\cL}{{\cal L}}
\newcommand{\cM}{{\cal M}}
\newcommand{\cN}{{\cal N}}
\newcommand{\cO}{{\cal O}}
\newcommand{\cQ}{{\cal Q}}
\newcommand{\cS}{{\cal S}}
\newcommand{\cY}{{\cal Y}}
\newcommand{\cU}{{\cal U}}
\newcommand{\cV}{{\cal V}}
\newcommand{\cW}{{\cal W}}
\newcommand{\cZ}{{\cal Z}}
\newcommand{\mkH}{\mathfrak{H}}
\newcommand{\tcA}{\widetilde{\cal A}}
\newcommand{\DD}{{\cal D}}
\newcommand\TYPE[3]{ \underset {(#1)}{\overset{{#3}}{#2}}  }
\newcommand{\Qc}{\overset{\footnotesize\circ}{Q}}
\newcommand{\bfe}{\boldsymbol e} 
\newcommand{\bfb}{{\boldsymbol b}}
\newcommand{\bfd}{{\boldsymbol d}}
\newcommand{\bfh}{{\boldsymbol h}}
\newcommand{\bfj}{{\boldsymbol j}}
\newcommand{\bfn}{{\boldsymbol n}}
\newcommand{\bfA}{{\boldsymbol A}}
\newcommand{\bfB}{{\boldsymbol B}}
\newcommand{\bfJ}{{\boldsymbol J}}
\newcommand{\bfS}{{\boldsymbol S}}
\newcommand{\dr}{\mathrm{d}}
\newcommand{\Dr}{\mathrm{D}}
\newcommand{\saddle}{\mathrm{saddle}}
\newcommand{\can}{\mathrm{can}}
\newcommand{\wt}[1]{\widetilde{#1}}
\newcommand{\wh}[1]{\widehat{#1}}
\newcommand{\ch}[1]{\check{#1}}
\newcommand{\ol}[1]{\overline{#1}}
\newcommand{\ii}{\imath}
\newcommand{\ic}{\iota}
\newcommand{\mbbP}{\mathbb{P}}
\newcommand{\mbbR}{\mathbb{R}}
\newcommand{\mbbN}{\mathbb{N}}
\newcommand{\mbbZ}{\mathbb{Z}}
\newcommand{\Leftrightup}[1]{\overset{\mathrm{#1}}{\Longleftrightarrow}}
\newcommand{\leftup}[1]{\overset{#1}{\longleftarrow}}
\newcommand{\avg}[1]{\left\langle\,{#1}\, \right\rangle}
\newcommand{\step}{\lrcorner\hspace*{-0.55mm}\ulcorner}
\newcommand{\GamLamM}[1]{{\Gamma\Lambda^{{#1}}\cal{M}}}
\newcommand{\GamLam}[2]{{\Gamma\Lambda^{{#2}}{#1}}}
\newcommand{\GTM}{{\Gamma T\cal{M}}}
\newcommand{\GT}[1]{{\Gamma T{#1}}}
\newcommand{\inp}[2]{\left\langle\,  #1\, , \, #2\, \right\rangle}
\title{
  Affine geometric description of thermodynamics
}
\author{ Shin-itiro GOTO 
  \\
  Center for Mathematical Science and Artificial Intelligence,\\
  Chubu University,\quad 
1200 Matsumoto-cho, Kasugai, Aichi 487-8501, Japan
}
%
\date{\today}
\maketitle
\begin{abstract}%
  Thermodynamics 
  provides a unified perspective of thermodynamic properties of various
  substances.
  To formulate thermodynamics in the language of sophisticated mathematics,  
  thermodynamics is described by a variety of  
  differential geometries, including contact and symplectic geometries.
  Meanwhile affine geometry is a branch of differential geometry and is
  compatible with information geometry,
  where information geometry is known to be
  compatible with thermodynamics.  
  By combining above, it is expected that thermodynamics is  compatible 
  with affine geometry, and is expected that
  several affine geometric tools can be introduced in the 
  analysis of thermodynamic systems.
  In this paper affine geometric descriptions of equilibrium and nonequilibrium thermodynamics are proposed. 
  For equilibrium systems, it is 
  shown that several thermodynamic quantities
  can be identified with geometric objects in affine geometry, and
  that several geometric objects can be introduced in thermodynamics. 
  Examples of these include: specific heat is identified with 
  the affine fundamental form, a flat connection is introduced in thermodynamic
  phase space. 
  For nonequilibrium systems, two classes of relaxation processes
  are shown to be described in the language of
  an extension of affine geometry.
Finally   
   this affine geometric description of thermodynamics
for equilibrium and nonequilibrium systems  
   is compared with 
  a contact geometric description.   
\end{abstract}%

%
\section{Introduction}
Thermodynamics is a branch of physics, provides a   
unified perspective of thermodynamic properties of various
substances, and has been applied to
various branches of sciences and technologies\,\cite{Callen}.
Thus further developments in thermodynamics are beneficial
in these branches. One way to develop thermodynamics further 
is to apply well-developed pure mathematics to thermodynamics\,\cite{Frenkel},
and differential geometry is one of such mathematics\,\cite{Nakahara}. 
By introducing notions developed in differential geometry to thermodynamics,
some new views and applications in thermodynamics were expected to be found.
Such views and applications are described by contact
geometry\,\cite{Harmann1973,Mrugala2000},  
symplectic geometry\,\cite{Baldiotti2016},
and so on\,\cite{Schaft2018entropy,Yoshimura2017}.
Here contact geometry is known as
an odd-dimensional counterpart of symplectic geometry
\,\cite{Arnold,Libermann1987,Silva2008,McInerney2013}, and 
is used in describing not only thermodynamics,
but also singularities in hyper-surfaces\,\cite{Arnold-Givental}.   
Note that there have been other geometric formulations
of thermodynamics\cite{Weinhold1976,Ruppeiner1995}, and various developments 
are in progress.

Affine geometry is a branch of differential geometry, and
studies invariant properties under affine transforms\,\cite{NomizuSasaki1994}. 
It is compatible with information geometry\,\cite{Matsuzoe2010},
where information geometry is a geometrization of mathematical 
statistics\,\cite{AmariNagaoka2000}. 
There are a variety of remarkable theorems relating 
affine geometry and 
information geometry. One of them
is that divergence that plays a central role in 
information geometry has been extended in the framework of affine
geometry\,\cite{Kurose1994}. 
Meanwhile information geometry is compatible with
thermodynamics\,\cite{Wada2015,Nakamura2019,Sagawa2022}.
Note that affine geometry is deeply related to the so-called Hessian
geometry\,\cite{Shima2007}.

By combining above, it is expected that thermodynamics is compatible with
affine geometry, and is expected that several geometric tools can be introduced
in the analysis of thermodynamic systems(see the diagram below):
$$
\xymatrix@C=40pt@R=20pt{
  \mbox{\fbox{Thermodynamics}}\ar@{.}[rr]|{\small\mbox{a potential link}}
  &
  &\mbox{\fbox{Affine geometry}}
  \\
  &
  \mbox{\fbox{Information geometry}}\ar@{-}[ul]|{\small\mbox{an existing link}}
  \ar@{-}[ur]|{\small\mbox{an existing link}}
  &
 }
$$

This paper is intended to
discuss relations between thermodynamics and affine geometry,  
and to provide the first step towards the materialization of
affine geometric descriptions of thermodynamics. 
To this end, an affine geometric description of thermodynamics 
and that of thermodynamic processes are proposed and discussed in this paper.
More specifically the following are shown: 
\begin{itemize}
\item
  The point of departure for this paper is to identify the set of
  equilibrium states in
thermodynamics with the image of a graph immersion 
(see Interpretation\,\ref{interpretation-basic} of this paper).     

\item
For equilibrium systems it is shown that several thermodynamic quantities
can be identified with geometric objects employed in affine geometry, and
that several geometric objects can be introduced in thermodynamics
(see Proposition\,\ref{fact-equilibrium-thermodynamics-correspondence-1} of this paper). 
Such examples include: specific heat is identified with 
affine fundamental form, a flat connection is introduced in thermodynamic
phase space.

\item
For nonequilibrium systems two classes of relaxation processes
are shown to be described in the language of an extension of 
affine geometry 
(see Theorems\,\ref{fact-induced-relaxation} and
\ref{fact-induced-relaxation-2} of this paper).
An analysis of a simple spin model with a unique set of equilibrium
states shows how the description is justified. 

\item
This affine geometric description of thermodynamics 
for equilibrium and nonequilibrium systems 
is compared with  
a contact geometric one, where the contact geometric thermodynamics is a representative existing theory.  
It is then shown that the present geometric formulation of
relaxation processes are consistent with the existing theory
(see Theorem\,\ref{fact-comparison-vector-fields} of this paper).
\end{itemize}

The rest of this paper is organized as follows.
In Section\,\ref{section-preliminaries}, some of necessary background of 
geometries and thermodynamics are summarized. 
In Section\,\ref{section-geometric-description-thermodynamics} an
affine geometric thermodynamics is proposed.
In Section\,\ref{section-comparison},
the affine geometric description of thermodynamics is compared
with the
contact geometric thermodynamics. 
Finally Section\,\ref{section-Conclusions} summarizes the present
geometric formulation of equilibrium and nonequilibrium thermodynamics,
and discusses future studies.

\section{Preliminaries}
\label{section-preliminaries}

This section is intended to provide a  
brief summary of the necessary background of 
geometries and thermodynamics, and is intended to fix notations here.  
Throughout this paper manifolds are connected, and 
every object on any manifold is smooth, 
unless otherwise stated. Given a manifold $\cM$, its tangent and cotangent
bundles are denoted by $T\cM$ and $T^{*}\cM$, respectively.  
Various formulae and 
tools developed in differential geometry are known\,\cite{Frenkel,Nakahara}.  
For example, the Lie derivative of a $k$-form $\alpha\in\GamLamM{k}$ on
a manifold $\cM$ 
along a vector field $X\in\GTM$ can be written as
$\cL_{X}\alpha=\dr\ii_{\,X}\alpha+\ii_{\,X}\dr \alpha$, where 
$\dr$ is the exterior derivative and $\ii_{\,X}$ the interior product with
$X$. This is known as the Cartan formula. This formula is valid even $k=0$,
where $\GamLamM{0}$ is identified with the space of functions on $\cM$.
If $\phi$ is a map from a manifold to another one,
then $\phi_{*}$ denotes the push-forward of $\phi$, and $\phi^{*}$ denotes
its pull-back. 
When a (affine) connection $\nabla$ is equipped on $\cM$,   
the $(1,2)$-type tensor field $T^{\nabla}$ such that  
$T^{\nabla}(X,Y):=\nabla_XY-\nabla_YX-[X,Y]$ for all $X,Y\in\GTM$, 
is called torsion tensor field, where $[X,Y]=XY-YX$ is the Lie bracket.
If $T^{\nabla}(X,Y)\equiv 0$ for all $X,Y\in\GTM$, then 
the $\nabla$ is said to be torsion-free.
The $(1,3)$-type tensor field $R^{\nabla}$ such that
$R^{\nabla}(X,Y)Z=\nabla_{X}\nabla_{Y}Z-\nabla_{Y}\nabla_{X}Z-\nabla_{[X,Y]}Z$ for all vector fields $X,Y,Z$, is called the curvature tensor field.
If $R^{\nabla}(X,Y)Z\equiv 0$, then $\nabla$ is called curvature-free. 
If a connection $\nabla$ is torsion-free and curvature-free, then
$\nabla$ is said to be {\it flat}.  If a connection $\nabla$ is flat, then
there exists a coordinate system such that
all the connection coefficients vanish, $\Gamma_{ij}^{\ \ k}\equiv 0$.
These coordinate systems are said to be affine coordinate systems. 

\subsection{Affine geometry}
\label{section-preliminary-Affine-geometry}

In this subsection necessary background in
affine geometry are summarized
(see Ref.\,\cite{NomizuSasaki1994} for more details. 
In Ref.\,\cite{NomizuSasaki1994},
(co)tangent space at a point is identified with a point due to
the property of affine space. 
Meanwhile in this paper, this identification is not adopted.).
Let $\cM$ be an $n$-dimensional manifold ($n=1,2,\ldots$), 
$f$ an immersion of $\cM$ into $\mbbR^{n+1}$, and
$\xi$ a vector field along $f$. 
If for arbitrary point $p\in \cM$ the condition 
$$
T_{f(p)}\mbbR^{n+1}
=f_{*}(T_{p}\cM)\oplus \mathrm{span}\,\{\xi_{p}\},
$$
is satisfied, then the pair $(f,\xi)$ is referred to as an
{\it affine (hyper-surface) immersion},
and $\xi$ a {\it transversal vector field}.

There are various formulae associated with affine immersions. 
Let $(f,\xi)$ be an affine 
hyper-surface immersion, and 
$\Dr$ the standard 
flat affine connection on $\mbbR^{n+1}$.    
Then
there exists a torsion-free connection $\nabla$ on $\cM$ such that
\beq
\Dr_X (f_{*}Y)
=f_{*}(\nabla_X Y)+h(X,Y)\,\xi,\qquad\forall\ X,Y\in T_{p}\cM
\label{Gauss-formula}
\eeq
where $h$ is a $(0,2)$-type symmetric tensor field called an  
{\it affine fundamental form} associated with $\xi$. 
Equation\,\fr{Gauss-formula}
is known as the Gauss formula\cite{NomizuSasaki1994}.
Throughout this paper,
immersions are always affine hyper-surface immersions unless otherwise stated. 
For all $X\in \GTM$, 
one can write the Weingarten formula: 
\beq
\Dr_X\xi
=-f_{*}(SX)+\tau(X)\,\xi,
\label{Weingarten-formula}
\eeq
where $S$ is a $(1,1)$-type tensor field being referred to as 
an {\it affine shape operator}, and $\tau$ a one-form being referred to as
a {\it transversal connection form}.
When $\tau=0$, the affine immersion $(f,\xi)$ is
referred to as being {\it equiaffine}.
If $h$ is non-degenerate everywhere, 
then $f$ is said to be {\it non-degenerate}.

The dual of vector spaces provides various tools
in differential geometry in general. 
In affine geometry, the dual of a vector space 
also provides useful geometric tools. 
Given an affine immersion $(f,\xi)$, 
introduce a map $v:\cM\to T_{f(-)}^{*}\mbbR^{n+1}$, and
the pairing
$\inp{-(p)}{-_{p}}:T_{f(p)}^{*}\mbbR^{n+1}\times T_{f(p)}\mbbR^{n+1}\to\mbbR$,
$(p\in\cM)$:
$$
\xymatrix{
  \mbbR^{n+1}&\cM\ar[l]^(0.3){f}\ar[r]_(0.3){v}&T_{f(-)}^{*}\mbbR^{n+1}
  \ar@{<.>}^{dual}[rr]
  &&T_{f(-)}\mbbR^{n+1}&T\cM\ar[l]^(0.36){f_{*}}\ar[r]_(0.3){v_{*}}
   &T(T_{f(-)}^{*}\mbbR^{n+1}).
}
$$
If a map $v$ satisfies the conditions 
\beq
\inp{v(p)}{\xi_{p}}
=1,\qquad\mbox{and}\qquad
\inp{v(p)}{f_{*}X_{p}}
=0,\qquad\quad\forall X_{p}\in T_{p}\cM,\quad \forall\ p\in\cM,
\label{conormal-map-conditions}
\eeq
then $v$ is referred to as the {\it conormal map}. 

The graph immersion is a class of affine immersions. 
  This is explained below. 
Let $\Omega$ be an $n$-dimensional region  
of $\mbbR^{n}$, and 
$F$ a function defined on
$\Omega$, $F:\Omega\to\mbbR$.
The coordinate system of $\Omega\subset\mbbR^{n}$
is denoted by $x=(x^{\,1},\ldots,x^{n})$, and the coordinate of
the other $\mbbR$ is denoted by $z$. 
A {\it graph immersion} associated with $F$ is a pair
$(f,\xi)$ satisfying the conditions written in coordinates as 
$$
x=(x^{\,1},\ldots,x^{n})\mapsto
f(x)
=(x,F(x)),\quad
\mbox{  and}\quad
\xi=\frac{\partial}{\partial z}.
$$
It follows from the flatness of $\Dr$ that $\Dr_X\xi=0$ for all 
$X\in T_{p}\cM$.
This and \fr{Weingarten-formula} yield 
$S=0$ and $\tau =0$. Since $\tau=0$,  
any graph immersion is equiaffine.
The affine fundamental form $h$ is calculated as follows. 
First, one calculates 
\beq
f_{*}\frac{\partial}{\partial x^{1}}
=\frac{\partial}{\partial x^{1}}
+\frac{\partial F}{\partial x^{1}}\frac{\partial}{\partial z},
 \quad\cdots,\quad
f_{*}\frac{\partial}{\partial x^{n}}
=\frac{\partial}{\partial x^{n}}
+\frac{\partial F}{\partial x^{n}}\frac{\partial}{\partial z},
\label{pushforward-graph-immersion-d-dx}
\eeq
From \fr{pushforward-graph-immersion-d-dx}, one calculates  
\beq
\Dr_{\frac{\partial}{\partial x^{i}}}f_{*}\frac{\partial}{\partial x^{j}}
=\frac{\partial^2 F}{\partial x^{i}\partial x^{j}}\frac{\partial}{\partial z},
\qquad i,j=1,\ldots,n.
\label{D-pushforward-graph-immersion-d-dx}
\eeq
From \fr{Gauss-formula} and \fr{D-pushforward-graph-immersion-d-dx}, one has 
\beq
h\left(\frac{\partial}{\partial x^{i}},\frac{\partial}{\partial x^{j}}\right)
=\frac{\partial^2 F}{\partial x^{i}\partial x^{j}},
\qquad i,j=1,\ldots,n,
\label{affine-fundamental-form-F}
\eeq
and
\beq
\nabla_{\frac{\partial}{\partial x^{i}}}\frac{\partial}{\partial x^{j}}
=0,\qquad i,j=1,\ldots, n.
\label{affine-graph-induced-connection}
\eeq
It follows from \fr{affine-fundamental-form-F}
 that if the matrix $(\partial^2 F/\partial x^{i}\partial x^{j})$ is strictly
positive at $x$, then $h$ is non-degenerate. In addition, it follows 
      from \fr{affine-graph-induced-connection} that the coordinate 
      system $x$ is affine with respect to $\nabla$.
      If  $h$ is non-degenerate everywhere on $\cM$, 
 then the graph immersion is said to be non-degenerate.  
The conormal map at $p\in\Omega$ is expressed as 
\beq
v(p)
=\dr z-\sum_{a=1}^{n}\frac{\partial F}{\partial x^{a}}\dr x^{a}\quad
\in\ T_{f(p)}^{*}\mbbR^{n+1}.
\label{conormal-map-graph-immersion}
\eeq
\subsection{Information geometry}
\label{section-preliminary-Information-geometry}

Information geometry is a geometrization of mathematical statistics\,
\cite{AmariNagaoka2000},
and a study of statistical manifolds\,\cite{Matsuzoe2010}.    
The definition of
statistical manifold is given from a viewpoint of differential geometry
as follows.  
Let $(\cM,g)$ be an $n$-dimensional (pseudo-) Riemannian manifold, and $\nabla$ a torsion-free
affine connection. If $\nabla$ satisfies the Codatti
equation\,\cite{NomizuSasaki1994},
$$
(\nabla_X g)(Y,-)
=(\nabla_Y g)(X,-),\qquad
\forall\,X,Y\in\GTM
$$
then the triplet $(\cM,\nabla,g)$ is referred to as a
{\it statistical manifold}\,\cite{Kurose1994}.

One pair of connections explained below plays
various roles in information geometry. Let $g$ be a pseudo-Riemannian  
metric tensor field on $\cM$, and $\nabla$ an affine connection. 
If a connection $\nabla^{*}$ satisfies the condition
$$
X[g(Y,Z)]
=g(\nabla_{X}^{*}Y,Z)+g(Y,\nabla_{X}Z),\qquad \forall\ X,Y,Z\in \GTM
$$
then $\nabla^{*}$ is referred to as the {\it dual connection} of $\nabla$
with respect to $g$. 
It can be shown that if $\nabla$ is flat then $\nabla^{*}$ is flat.  
In information geometry,
the tetrad $(\cM,g,\nabla,\nabla^{*})$ is referred to as
a {\it dually flat space}, and this class of manifolds has been
well-studied\,\cite{Matsuzoe2010}. 
On a dually flat space, let $\theta=(\theta^{1},\ldots,\theta^{n})$
be an affine coordinate system. Then it can be shown that there
exists an affine coordinate system
$\eta=(\eta_{1},\ldots,\eta_{n})$ for $\nabla^{*}$ that 
satisfies  
$$ 
g\left(
\frac{\partial}{\partial\theta^{a}},\frac{\partial}{\partial\eta_{b}}
\right)
=\delta_{a}^{b}.\qquad
a,b=1,\ldots,n,
$$
where $\delta_{a}^{b}$ is the Kronecker delta
giving unity when $a=b$ and zero when $a\neq b$. 
The coordinate system $\eta$ is said to be the
{\it dual coordinate system} of $\theta$ with respect to $g$. 
One basic proposition on  dually flat spaces is as follows.
Let $(\cM,g,\nabla,\nabla^{*})$ be a dually flat space, $\theta$
an affine coordinate system, $\eta$ a dual affine coordinate system, 
where
$\cM$ is simply connected and has a global coordinate system.
Then there exist functions $\psi$ and $\varphi$, and it follows that
\beqa
&&\frac{\partial\psi}{\partial \theta^{a}}
=\eta_{a},\quad
\frac{\partial\varphi}{\partial \eta_{a}}
=\theta^{a},\quad
\psi(p)+\varphi(p)-\sum_{a=1}^{n}\theta^{a}(p)\eta_{a}(p)
=0,\quad a,b=1,\ldots,n,\quad \forall\ p\in \cM
\label{Legendre-relations-1}\\
&&
g_{ab}
=\frac{\partial^2\psi}{\partial\theta^{a}\partial \theta^{b}},\quad
g^{ab}
=\frac{\partial^2\varphi}{\partial\eta_{a}\partial \eta_{b}},\quad
a,b=1,\ldots,n
\non
\eeqa
where $g_{ab}=g(\partial/\partial\theta^{a},\partial/\partial\theta^{b})$
and $(g^{ab})$ is the inverse matrix of $(g_{ab})$.  
In mathematical statistics and information geometry,  
divergence plays various roles\,\cite{AmariNagaoka2000}.
The function
$\cM\times\cM\to \mbbR$, 
$$
D(p_1,p_2)
=\psi(p_{1})+\varphi(p_{2})-\sum_{a=1}^{n}\theta^{a}(p_{1})\eta_{a}(p_{2}), 
$$
is called the {\it canonical divergence}. 

Several relations in affine geometry and information geometry are known
in the literature, and  one of them is as follows:
\begin{Proposition}
\label{fact-affine-statistical-manifold-1}
  (\cite{Matsuzoe2010}).
  If $(f,\xi)$ is non-degenerate and equiaffine, then $(\cM,\nabla,h)$
  is a statistical manifold.
\end{Proposition}
Another relation  
between affine and information geometries is found on
the study of divergence.
To define the geometric divergence, 
consider a graph immersion $(f,\xi)$ associated 
with $\psi:\Omega\to\mbbR$, written in coordinates as 
$$
f:\theta\mapsto (\theta,\psi(\theta)),\quad
\xi=\frac{\partial}{\partial z}.
$$
If $\psi$ is convex, 
then the explicit form of the conormal map acting on $p\in\Omega$ is
obtained in coordinates with \fr{conormal-map-conditions} and 
\fr{Legendre-relations-1} as 
\beq
v(p)
=\dr z-\sum_{a=1}^{n}\eta_{a}(p)\,\dr \theta^{a}\quad
\in\ T_{f(p)}^{*}\mbbR^{n+1}.
\label{conormal-map-at-p}
\eeq
Then, let $(f,\xi)$ be
a non-degenerate equiafffine immersion, 
 $v$ its conormal map, and 
$\Delta^{f}:\Omega\times\Omega\to T\mbbR^{n+1}$ a map such that 
$$
\Delta^{f}(p_{1},p_{2})
=\sum_{a=1}^{n}(\theta^{a}(p_{1})-\theta^{a}(p_{2}))
\frac{\partial}{\partial \theta^{a}}+
(\psi(p_{1})-\psi(p_{2}))\frac{\partial}{\partial z}\quad
\in T_{f(p_{2})}\mbbR^{n+1}.
$$
Then the function $D^{\G}:\Omega\times\Omega\to\mbbR$, called
{\it geometric divergence},   
\beq
D^{\G}(p_{1},p_{2})
=\inp{v(p_{2})}{\Delta^{f}(p_{1},p_{2})},
\label{geometric-divergence}
\eeq
is introduced in the context of the study of affine geometry. 
A relation between $D$ and $D^{\G}$ is obtained
as shown explicitly in Ref.\,\cite{Matsuzoe2010}. 
It is obtained from \fr{conormal-map-at-p} and 
$$
\varphi(p_{2})
=-\psi(p_{2})+\sum_{a=1}^{n}\eta_{a}(p_{2})\theta^{a}(p_{2}),
$$
as 
\beqa
D^{\G}(p_{1},p_{2})
&=&\inp{v(p_{2})}{\Delta^{f}(p_{1},p_{2})}
\non\\
&=&\psi(p_{1})-\psi(p_{2})
-\sum_{a=1}^{n}\eta_{a}(p_{2})(\theta^{a}(p_{1})-\theta^{a}(p_{2}))
\non\\
&=&\psi(p_{1})+\varphi(p_{2})-\sum_{a=1}^{n}\eta_{a}(p_{2})\theta^{a}(p_{1})
\non\\
&=&D(p_1,p_2).
\non
\eeqa

In more general case, a relation between the divergence and 
the geometric divergence has been known as follows: 
\begin{Theorem}
\label{fact-divergences-coincide}
  (\cite{Matsuzoe2010}). 
  Let $(\cM,\nabla,g)$ be a simply connected dually flat space. 
  Then the canonical divergence and the geometric divergence on
  $(\cM,\nabla,g)$ coincides.
\end{Theorem}

\subsection{Thermodynamics and existing geometric formulations  }
\label{section-preliminary-thermodynamics}
In this subsection necessary background of thermodynamics for this study
is briefly summarized. 
Symbols introduced here for thermodynamic quantities are duplicated with
symbols for geometry introduced
in Section\,\ref{section-preliminary-Affine-geometry}.  
These duplicated symbols become consistent when affine geometric thermodynamics
is constructed.
In addition, thermodynamic processes in this paper
are assumed to be quasi-static for simplicity. 

To formulate equilibrium thermodynamics, we employ 
a subset of $\mbbR^{n}$ for thermodynamic variables. 
In addition
we employ $\mbbR$ for 
a complete thermodynamic function, where
{\it complete thermodynamic functions}  
are functions that derive equations of state at equilibrium and response
functions\,\cite{Sasa2000}.
Examples of complete thermodynamic functions are the  
internal energy, entropy, 
Helmholtz and Gibbs free-energies with appropriate arguments. 
Thermodynamic variables and complete thermodynamic functions
play fundamental roles, 
in the sense that these induce equations of state at equilibrium. 
More details about these variables and functions are explained below. 
 Let $x=(x^{1},\ldots,x^{n})$ be a set of 
 thermodynamic variables
in $\mbbR^{n}$ and
$\cF$ a complete thermodynamic function
(free-energy or internal energy).
If $x$ is a set of arguments or equivalently variables of $\cF$,
  then  $x^{1},\ldots,x^{n}$ are called
{\it primal thermodynamic variables} in this paper. 
Then the set of the
{\it thermodynamic conjugate variables} 
$y=(y_{1},\ldots,y_{n})$ is such that
\beq
y_{a}
=\frac{\partial \cF}{\partial x^{a}},\quad
a=1,\ldots,n.
\label{conjugate-variables-general}
\eeq
In addition the value of a complete thermodynamic function $\cF$
(a free-energy, entropy,  or internal energy) 
at $x$ should be the
same as that of $\cF(x)$,
\beq
z=\cF(x)
\label{free-energy-value-equilibrium}.
\eeq
The thermodynamic phase space is where
\fr{conjugate-variables-general} and \fr{free-energy-value-equilibrium}
are  satisfied. 
Hence, the thermodynamic phase space is a subset of $\mbbR^{2n+1}$. 
The fundamental relation of thermodynamics can be written as
\beq
\dr U-T\,\dr S+P\,\dr V
=0,
\label{thermodynamic-fundamental-relation}
\eeq
where $U$ is internal energy, $T$  the absolute temperature, $S$ entropy,
$P$ pressure, and $V$ volume. In addition, 
the heat (one-form) $\cQ$ and the work (one-form) $\cW$ are defined as 
$$
\cQ
=T\,\dr S,\qquad\mbox{and}\qquad
\cW
=-P\,\dr V,
$$
respectively\,\cite{Frenkel}.
The first law of thermodynamics 
states that there exists the function $U$ so that
\fr{thermodynamic-fundamental-relation} holds with some processes,
where processes are integral curves of vector fields. 
In other words, it states that the sum $\cQ$+$\cW$ is
an exact one-form $\dr U$. 
Meanwhile, the second law of thermodynamics states that there exists
the function $S$ with some properties.
To change variables, the Legendre transform is applied to functions, 
where the transformed functions are convex if the original functions are
convex. 
Given a function $\cF$, it follows from the theory of Legendre transform
that there is a function
$\cF^{*}$ such that 
$$
x^{a}
=\frac{\partial \cF^{*}}{\partial y_{a}},
\qquad a=1,\ldots,n.
$$
After changing thermodynamic variables, 
\fr{thermodynamic-fundamental-relation}
can be written in various forms, such as
\beq
\dr S-\frac{1}{T}\dr U-\frac{P}{T}\dr V
=0,
\label{thermodynamic-fundamental-relation-S}  
\eeq
\beq
\dr \cA+S\dr T + P\dr V
=0,
\label{thermodynamic-fundamental-relation-A}
\eeq
and so on, where $\cA$ denotes the Helmholtz free-energy. 
Response functions, such as 
heat capacity and specific heat,
are obtained by differentiation of \fr{conjugate-variables-general} 
as 
\beq
\chi_{ab}
=\frac{\partial y_{a}}{\partial x^{b}}
=\frac{\partial^2 \cF}{\partial x^{a}\partial x^{b}},\qquad
a,b=1,\ldots,n.
\label{chi-a-b}
\eeq

Equilibrium states can be classified into at least three classes. 
They are unstable, metastable, and most stable equilibrium states.
Since equilibrium states are fundamental objects in the study of
thermodynamics, a classification of equilibrium states provides
further understanding of thermodynamic properties of substances.  
The three classes above are explained below.  
The most stable equilibrium states are equilibrium states
that are structurally stable within some time-length against
some small external perturbation.  
Metastable equilibrium states are states that exist without very small
external perturbation, but states deviate from the metastable states  
under some strength of external perturbation. 
Unstable equilibrium are states that are hard to be realized experimentally
since the states will not return to the original 
unstable equilibrium states under very small perturbation. 

Nonequilibrium thermodynamics is a developing branch of physics, and
several theories have been proposed
in the literature\cite{Kubo1991,Zubarev1996}. 
To provide a solid foundation for a nonequilibrium theory, 
we focus on a simple class for clarity. One simple class of nonequilibrium
phenomena is that of relaxation processes. In this paper
even in a nonequilibrium state, thermodynamic variables are assumed to be
defined and described by extending the equilibrium thermodynamic variables.
Let $y(t)$ and $\cF(t)$ be a set of thermodynamic variables and 
the value of a nonequilibrium free-energy at time $t\in I\subset \mbbR$,
respectively. 
In addition, $y$ and $\cF$ denote the corresponding
variable set and the function defined at equilibrium, respectively.  
If a process (or a time-evolution) that satisfies  
$$
\lim_{t\to \infty}y(t)
=y,\qquad\mbox{and}\qquad
\lim_{t\to \infty}\cF(t)
=\cF,
$$
then the process is referred to as a relaxation process in this paper.  

There are several existing geometric formulations of thermodynamics
in the literature.
Such existing studies include Hesse geometry and contact geometry: 
\begin{itemize}
\item
(\cite{Shima2007}). 
  A Hesse manifold is a manifold $\cM$
  equipped with a structure $(\Dr,g)$, where   $\Dr$ is a flat connection,
  and $g$ a pseudo Riemannian metric tensor field that can be written as 
  $g=\Dr\dr \cF$ with $\cF$ being a function.  

\item
   (\cite{Silva2008,Libermann1987}).
  A contact manifold is an odd-dimensional manifold $\cM$ equipped with 
  a distribution $\ker\lambda$, where $\lambda$ is a one-form that satisfies  
  the condition that the top-form
  $\lambda\wedge\dr\lambda\wedge \cdots\wedge\dr\lambda$ is a 
  volume element on $\cM$. 
\end{itemize}  

In this section the calligraphic letter $\cF$ is used to emphasize 
that the function is a complete thermodynamic function. 
In what follows this emphasis is not adopted, and thus the calligraphic letter
is not used even in the case that a function is a complete thermodynamic function.

\section{Affine geometric description of thermodynamics}
\label{section-geometric-description-thermodynamics}
In this section an affine geometric description of thermodynamics is 
proposed and discussed. 
This description consists of two cases, one is
equilibrium case and the other nonequilibrium one.
The description of the nonequilibrium case can be divided into two, one is
the case of a unique set of  equilibrium states,
and the other is the case of   
two sets of equilibrium states:
\beqa
&\bullet&\mbox{Equilibrium}
\non\\
&\bullet&\mbox{Nonequilibrium}
\non\\
&&\left\{\begin{array}{l}
\mbox{a unique set of equilibrium states}\\
\mbox{two sets of equilibrium states}
\end{array}
\right.\ .
\non
\eeqa

\subsection{Equilibrium}
\label{section-geometric-equilibrium}
To describe thermodynamics in the language of 
affine geometry,
the basic idea proposed in this paper is to employ graph immersions of 
a region $\Omega\subset\mbbR^{n}$ into $\mbbR^{n+1}$.
In this paper the following interpretations are
proposed.
\begin{Interpretation}
\label{interpretation-basic}
  (Equilibrium thermodynamics and affine geometry). 
  \begin{enumerate}
\item
  A point of $\mbbR^{n+1}$ is identified as  
  a set of  primal thermodynamic variables 
  and the value of 
  a complete thermodynamic function 
  (a free-energy, entropy or internal energy).
  
\item
  The coordinate system $x=(x^{1},\ldots,x^{n})$ for $\Omega\subset\mbbR^{n}$
  represents primal thermodynamic variables
  in thermodynamic systems. 
  Then the coordinate $z$ for the additional space $\mbbR$ represents   
  the value of a complete thermodynamic function
  (a free-energy, entropy, or internal energy). This is written as
  $z=F(x)$ when a thermodynamic relation
  holds, and let $\xi=\partial/\partial z$. 

\item
  The image of a graph immersion $(f,\xi)\subset\mbbR^{n+1}$
  associated with $F$  
  is identified with a thermodynamic phase space, where
  a thermodynamic relation holds. Accordingly, a point of this
  space is identified with an equilibrium state.

\item
  The conormal map is identified with one of the expressions of the
  fundamental relation of thermodynamics, 
  \fr{thermodynamic-fundamental-relation},
  \fr{thermodynamic-fundamental-relation-S},
  and \fr{thermodynamic-fundamental-relation-A}. 

\end{enumerate}
\end{Interpretation}
Note that as well as item 1 of Interpretation\,\ref{interpretation-basic}
where $(n+1)$-dimensional manifolds are involved, 
there are several geometric approaches with $(n+1)$-dimensional
manifolds in the literature\,\cite{Frenkel,Mrugala1978}.

\begin{Remark}
\label{fact-equilibrium-thermodynamics-correspondence-0}
  The above identifications induce 
the following.

\begin{enumerate}
\item 
  The affine fundamental form $h$, a $(0,2)$-type operator, is identified
  with a set of response functions at equilibrium
  (on the thermodynamic phase space) 
  (see \fr{affine-fundamental-form-F} and
  \fr{chi-a-b}). 

\item 
  The flat connection $\nabla$ on $\Omega$ is introduced  
  at equilibrium (on the thermodynamic phase space). 
  In addition,  $x$ is an affine coordinate system  
  (see \fr{Gauss-formula} and \fr{affine-graph-induced-connection}). 

\item 
  The $z$ component of the push-forward $f_{*}(\partial/\partial x^{a})$
  is identified with the thermodynamic conjugate variable with respect to
  $x^{a}$ (see \fr{pushforward-graph-immersion-d-dx}).
  This is written as $y_{a}=\partial F/\partial x^{a}$, where
  $y_{a}$ is the thermodynamic conjugate variable with respect to
  a primal thermodynamic variable $x^{a}$, and the collection is denoted by
  $y=(y_{1},\ldots,y_{n})$.
  Hence thermodynamic primal and conjugate variables
  are expressed as a point of the tangent bundle $T(f\Omega)$.    

\item 
  A succinct information about describing a set of
  equilibrium states is the image of  
  a graph immersion $(f,\xi)$ into $\mbbR^{n+1}$, so that
  the definition of equilibrium state is given by item 3 of
  Interpretation\,\ref{interpretation-basic}.
  When thermodynamic conjugate variables are needed, its tangent bundle is
  considered. 

  For later purpose of describing nonequilibrium states, 
  a relaxation (or, an extension) of the image of the
  graph immersion is discussed here. 
  Since the values of a free-energy and thermodynamic variables
  in a nonequilibrium state
  are not written as $F(x)$ and $(x,y(x))$ in coordinates,  
  the manifold $T(f\Omega)$ is relaxed to (or, extended to)
\beq
  T\Omega\times\mbbR.
  \label{projection-Tf}
  \eeq
  A point of
  $T\Omega\times\mbbR$ can be written in coordinates as $(x,y,z)$,
  where $(x,y)$ is the coordinates of $T\Omega$ and
  $z$ is the coordinate of $\mbbR$.  
  Meanwhile the manifold $T\Omega\times\mbbR$ is redundant for
  the purpose of expressing equilibrium states.
Since $T(f\Omega)$ and $T\Omega\times\mbbR$ include redundant components
for describing equilibrium states,
we have defined equilibrium states as the image of 
a graph immersion $(f,\xi)$. 

\item
  The conormal map, \fr{conormal-map-graph-immersion} and
  \fr{conormal-map-conditions},  
  provides 
  the fundamental relation of thermodynamics,
  \fr{thermodynamic-fundamental-relation}
    , \fr{thermodynamic-fundamental-relation-S},
  and \fr{thermodynamic-fundamental-relation-A}.   
  
\item
  In the case that $F$ is convex for the graph immersion $(f,\xi)$,
  it follows from Proposition\,\ref{fact-affine-statistical-manifold-1}
  that the equilibrium phase space is identified with a statistical manifold.

\item
  In the case that $F$ is convex, 
  the geometric divergence \fr{geometric-divergence} 
  is introduced in the thermodynamic phase space. 
\end{enumerate} 
\end{Remark}

Some of Remark\,\ref{fact-equilibrium-thermodynamics-correspondence-0}
is summarized as follows. 
This is the main claim in this subsection:  
\begin{Proposition}
\label{fact-equilibrium-thermodynamics-correspondence-1}
(equilibrium states in the language of affine geometry).    
On a set of equilibrium states that is 
  a thermodynamic phase space 
in the sense of this paper, 
an affine fundamental form and a flat connection are induced.
In addition,
thermodynamic conjugate variables are described in the tangent bundle.
The  geometric divergence
is introduced if the complete thermodynamic function
is convex or concave.
\end{Proposition}

Examples are given as follows. 
The first example below shows 
  how to find an affine immersion and its geometric quantities 
  from a given equation of state and a given complete thermodynamic function. 
\begin{Example}
\label{example-ideal-gas-1}  
  (Ideal gas\ and its Helmholtz free energy).
  Consider the ideal gas, where the equation of state
  is
  $$
  PV=RT,
  $$
  where $R>0$ is constant, $T>0$ temperature, $P$ pressure, and $V$ volume.
This equation is written as 
$$
P=-\frac{\partial \cA}{\partial V},\qquad
\cA=-RT\ln V,
$$
where $\cA$ 
denotes the Helmholtz free-energy. Let $T$ be fixed, and identify
$\Omega=\mbbR_{>0}$,
$$
x=V,\qquad
F(x)=-\cA(V)
=RT\ln x,
$$
and $z$ the value of the free-energy.
  From the Helmholtz free-energy as a complete thermodynamic function and the
  equation of state given above, the corresponding affine immersion is
  shown below. As a transversal vector field, take $\xi=\partial/\partial z$.  
Then the image of the graph immersion $(f,\xi)$ associated with $F$ 
is identified with the set of equilibrium states
at temperature $T$. 
The affine fundamental form $h$ is such that
$$
h=\frac{\partial^2 F}{\partial x^2}\,\dr x\otimes\dr x,\qquad
\frac{\dr^2 F}{\dr x^2}
=-\frac{RT}{x^2}\qquad
\left(=\frac{\partial P}{\partial V}\right).
$$
Thus $h$ is non-degenerate on $\Omega=\mbbR_{>0}$, from which
this graph immersion is non-degenerate.
By applying Proposition\,\ref{fact-affine-statistical-manifold-1}
one can introduce the dual coordinate system $x^{*}$ with respect to $h$,
$$
F^{*}(x^{*})
=RT-RT\ln(RT)+RT\ln(x^{*}),
$$
and the geometric divergence $D^{\G}$. 
The $z$ component of  $f_{*}(\partial/\partial x)$, $y$,
 is identified with $P$.
\end{Example}

  The example below shows how to find an affine immersion and its geometric
  quantities from a given complete thermodynamic function and a given
  fundamental relation of thermodynamics. 
\begin{Example}
\label{example-ideal-gas-2}    
  (Ideal gas\ and its  entropy).
  Consider the ideal gas again (see Example\,\ref{example-ideal-gas-1}).
Introduce entropy $S$ as a complete thermodynamic function 
$$
S(U,V)
=R\ln (U^c V),
$$
where $U$ is internal energy and $c$ a positive constant. 
Since primal thermodynamic variables are
the arguments of $S$, the primal thermodynamic variables
defined on $\Omega=\mbbR_{>0}^{2}:=\mbbR_{>0}\times\mbbR_{>0}$
for this model are 
$$
x^{1}=U,\qquad\mbox{and}\qquad
x^{2}=V,
$$
so that $S:\Omega\to\mbbR$,
$$
S(x^{1},x^{2})
=R\ln(\,(x^{1})^{\,c} x^{2}\,).
$$
To specify an immersion, we 
let $z$ be the value of $S$ and let $\xi=\partial/\partial z$
be a transversal vector field.  
The image of the graph immersion associated with $S$ is
identified with the thermodynamic phase space. 
The conormal map is expressed and specified as 
$$
v=\dr z
-\frac{\partial S}{\partial x^{1}}\dr x^{1}
-\frac{\partial S}{\partial x^{2}}\dr x^{2}
=\dr S-\frac{1}{T}\dr U-\frac{P}{T}\dr V.
$$
The conjugate thermodynamic variables are then 
$$
y_{1}=\frac{\partial S}{\partial x^{1}}
=\frac{cR}{x^{1}}
=\frac{1}{T},\qquad\mbox{and}\qquad
y_{2}=\frac{\partial S}{\partial x^{2}}
=\frac{R}{V}
=\frac{P}{T}.
$$
These yield the expression of $U$ and the equation of state, 
$$
U=cRT,\qquad\mbox{and}\qquad
PV=RT,
$$
respectively. From the first equation above,
the specific heat $C=\dr U/\dr T$  is derived as $C=cR$.   
The explicit form of the affine fundamental form $h$ is shown as 
$$
h=\sum_{a=1}^{2}\sum_{b=1}^{2}
h_{ab}\,\dr x^{a}\otimes\dr x^{b},
\quad
h_{ab}
:=\frac{\partial^2 S}{\partial x^{a}\partial x^{b}},\quad\mbox{with}\quad 
h_{11}
=-\frac{cR}{(x^{1})^{2}},\quad
h_{12}
=0,\quad
h_{22}
=-\frac{R}{(x^{2})^{2}}.
$$
Thus $h$ is non-degenerate on $\Omega=\mbbR_{>0}^{2}$, from which
this graph immersion is non-degenerate. 
By applying Proposition\,\ref{fact-affine-statistical-manifold-1}
one can introduce the function $S^{*}$ so that
$$
x^{1}=\frac{\partial S^{*}}{\partial y_{1}},\quad\mbox{and}\quad
x^{2}=\frac{\partial S^{*}}{\partial y_{2}}.
$$
This $S^{*}(y_{1},y_{2})$ is found to be 
$$
S^{*}(y_{1},y_{2})
=R\ln (y_{1}^{\,c}y_{2})+\mbox{(constant)},
$$
and then the geometric divergence $D^{\G}$ can be introduced. 
\end{Example}

  There are several complete thermodynamic functions for the ideal gas system,
  such as $\cA$ 
  in Example\,\ref{example-ideal-gas-1} and $S$ 
  in Example\,\ref{example-ideal-gas-2}. Internal energy as a function of 
  $S$ and $V$ is another choice as a complete thermodynamic function.
  In this choice, 
  $U(S,V)=c^{\,\prime}V^{-1/c}\exp(S/(cR))$ and 
  \fr{thermodynamic-fundamental-relation} yield
  the equation of state $PV=RT$ and the specific heat $C=cR$,
  where $c^{\,\prime}$ is constant.  
  In the examples above, the affine immersions are non-degenerate.
  Meanwhile in the following example, an affine immersion is shown to be
  degenerate.
\begin{Example}
(van der Waals equation of state). 
  The state equation for the van der Waals gas model
  in the dimension-less variables 
is written as
$$
\left(P+\frac{3}{V^2}\right)(3V-1)
=8T,
$$
where $T$ denotes temperature, $V$ volume, and $P$ pressure. 
This equation can be written as 
$$
P=-\frac{\partial \cA}{\partial V},\qquad
\cA
=-\frac{3}{V}-\frac{8T}{3}\ln(3V-1), 
$$
where $\cA$ denotes the Helmholtz free-energy. 
Let $T$ be fixed, and identify $\Omega=\mbbR_{>0}$, 
$$
x=V,\quad
F(x)=-\cA(V),
$$
and $z$ the value of the free-energy.
Choose $\xi=\partial/\partial z$ as a transversal vector field. 
Then the image of the graph immersion $(f,\xi)$ associated with $F$ 
is identified with the set of equilibrium states
at temperature $T$.
The affine fundamental form $h$ is such that
$$
h=\frac{\partial^2 F}{\partial x^2}\,\dr x\otimes\dr x,\qquad
\frac{\partial^2 F}{\partial x^2}
=\frac{6}{x^3}-\frac{24 T}{(3x-1)^2}
\quad
\left(=\frac{\partial P}{\partial V}\right),
$$
from which this graph immersion is degenerate (not non-degenerate). 
The $z$ component of  $f_{*}(\partial/\partial x)$, $y$,
is identified with $P$. 
Note that there are several existing studies in the literature 
on this model
in the language of Riemannian geometry with the Levi-Civita
connection\,\cite{Ruppeiner1995,Brody1995}. 
\end{Example}

\subsection{Nonequilibrium}
\label{section-nonequilibrium}
In this subsection, a nonequilibrium geometric theory is proposed.
This is based on the equilibrium affine geometric theory developed in
Section\,\ref{section-geometric-equilibrium}.

To construct a nonequilibrium geometric theory 
one needs to
\begin{enumerate}
\item
  introduce physical time, 
\item
  extend or relax the equilibrium theory, where the equilibrium theory is  
based on a graph immersion
  $(f,\xi)$ into $\mbbR^{n+1}$, 
\item
verify that the nonequilibrium theory is consistent
with the equilibrium theory in some limits. 
One of these limits is the time-asymptotic limit.
\end{enumerate}

To develop a nonequilibrium theory, we 
\begin{enumerate}
\item
introduce time, and it is denoted by $t\in I\subset \mbbR$, 
\item
  introduce the trivial fiber bundle $\Omega\times\mbbR$, where 
  the restriction of this bundle to the set of equilibrium states is 
  $\Omega\times F(\Omega)$ being  
  the image of a graph immersion $(f,\xi)$.
  Hence this trivial bundle is an extension of the graph immersion. 
\item
  focus on the so-called relaxation process.
Relaxation processes are time-dependent phenomena in which
thermodynamic variables achieve an equilibrium state
from a nonequilibrium state through a time evolution.
Since there are a variety of classes of nonequilibrium systems and 
any confusion should be avoided, 
a simple formulation for a simple nonequilibrium phenomena 
is proposed in this paper. 
Then, 
time-asymptotic limits of the nonequilibrium theory is verified to be 
consistent with the equilibrium theory.
\end{enumerate}

\begin{figure}[htb]
  \includegraphics[width=8.4cm]{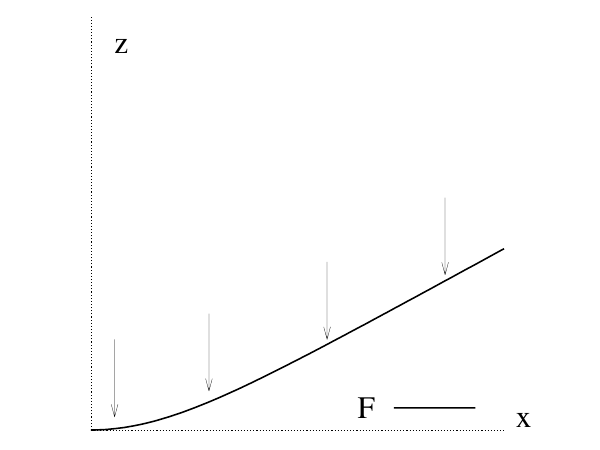}
  \hspace*{-14mm}
\includegraphics[width=8.4cm]{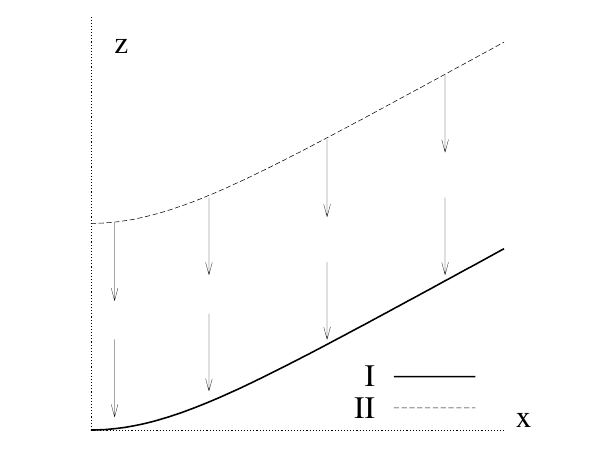}
\caption{ Relaxation processes as integral curves of vector fields.
  (Left) System with a unique set of equilibrium states.
  The thick line represents
  the set of unique equilibrium states that is the graph of $F$, and 
  the arrows represent the relaxation generating vector field $X_{F}$
  (see Section\,\ref{section-nonequilibrium-unique-equilibrium}).  
  (Right) System with two sets of equilibrium states. 
  The thick line represents the set of the
  most stable equilibrium state that is the graph of $F_{\I}$,
  the dashed line represents  
  the set of metastable equilibrium states 
  that is the graph of $F_{\II}$,  and 
  the arrows represent the relaxation generating vector field $X_{\II\to\I}$
  (see Section\,\ref{section-nonequilibrium-non-unique-equilibrium}).
} 
\label{picture-relaxation-vector-fields}
\end{figure}

  In what follows, the two cases are considered
(see Fig.\,\ref{picture-relaxation-vector-fields}).
  They are relaxation processes for systems with
  \begin{itemize}
\item     
a unique set of equilibrium states, and
\item
  two sets of equilibrium states.  
\end{itemize}

  \vspace*{-3mm}

\subsubsection{Nonequilibrium, (i) case of unique set of equilibrium states}
\label{section-nonequilibrium-unique-equilibrium}

In the following, a nonequilibrium geometric theory is proposed for the case
that there is a unique set of equilibrium states.   

The following is the definition of relaxation process in the case that
there is a unique set of equilibrium states.  

\begin{Definition}
\label{Definition-relaxation-process}
(relaxation process). Consider a graph immersion $(f,\xi)$ associated with
$F$ from $\Omega$ into  $\mbbR^{n+1}$, ($\dim\Omega=n$),
and let $x$ be a coordinate system of $\Omega$, 
$z\in\mbbR$, $\xi=\partial/\partial z$ and 
$y_{a}=\dr z(f_{*}(\partial/\partial x^{a}))$, 
($a=1,\ldots,n$). Note that $y_{a}$ is the $z$-component of
$f_{*}(\partial/\partial x^{a})$.  
Suppose that the image of $(f,\xi)$ is identified with
a set of equilibrium states.  
Then let $\phi:I\ni t\mapsto p(t)\in T\mbbR^{n+1}$ be a curve, 
where the coordinates of $p(t)$ are denoted by 
$(x(t),z(t),y(t),w(t))$, ($t\in I\subset \mbbR$).
For each $x$ kept fixed in $t$, if a curve satisfies the conditions    
\beq
\lim_{t\to\infty}y_{a}(t)
=\frac{\partial F}{\partial x^{a}}(x),\quad\mbox{and}\quad
\lim_{t\to\infty}z(t)
=F(x),\qquad
a=1,\ldots,n,
\label{conditions-relaxation-process}
\eeq
then the image of the  curve
is said to be a {\it relaxation process} 
towards a point of the set of 
equilibrium states $(f,\xi)$. 
\end{Definition}

To state an affine geometric description of
this class of relaxation processes in terms of vector fields,
a set of equilibrium states is placed as follows. 
Let $(f,\xi)$ be a graph immersion into $\mbbR^{n+1}$ 
written in coordinates as $x\mapsto (x,F(x))\in\mbbR^{n+1}$ and
  $\xi=\partial/\partial z$,    
that is, this immersion is associated with a function $F:\Omega\to\mbbR$.
First, recognize that $\Omega\times \mbbR\subset\mbbR^{n+1}$
is a trivial fiber bundle,
$\pi:\Omega\times\mbbR\to\Omega, \pi(q,z)=q$.
Hence $z\in\mbbR$ is a point of the fiber whose base point is $q$.   
Then, consider the vector field on $\mbbR$ over the point
$q\in\Omega$ of the form written in coordinates as 
\beq
X_{F}=w_{\,F}(z;x)\frac{\partial}{\partial z}
\label{XF}
\eeq
with $w_{\,F}(-;x)$ being a function of $z$ at a given $x$ 
(see Fig.\,\ref{picture-relaxation-vector-fields} (Left)). 
Integral curves of \fr{XF} are obtained by solving the ordinary differential
equation (ODE) 
\beq
\frac{\dr z}{\dr t}
=w_{\,F}(z;x),\qquad t\in I\subset \mbbR.  
\label{XF-ODE}
\eeq
Equation \fr{XF-ODE} is said to be the
{\it system of the ODE associated with $X_{F}$} 
in this paper. 
If a solution $z(t;x)$ of \fr{XF-ODE} satisfies 
$$
\lim_{t\to\infty}z(t;x)
=F(x),
$$
then the system \fr{XF-ODE} together with an appropriate
initial condition is referred to as a  
{\it relaxation generating system},  
and the corresponding vector field $X_{F}$ a 
{\it relaxation generating vector field} in this paper. 

The system  in the following example is 
a relaxation generating system.  

\begin{Example}
\label{Example-zeta=F-z}
Consider the dynamical system on $\mbbR$ associated with $X_{F}$: 
\beq
\frac{\dr z}{\dr t}
=w_{\,F}(z;x),\qquad
w_{\,F}(z;x)
=F(x)-z,\qquad \mbox{for each fixed $x$},\quad I=\mbbR. 
\label{ODE-z-F}
\eeq
Then  the explicit form of the solution $z(t;x)$
is immediately obtained from \fr{ODE-z-F} as 
\beq
z(t;x)
=(1-\e^{\,-t})F(x)+\e^{\,-t} z(0;x)
=:F_t(x).
\label{ODE-z-F-solution}
\eeq
From $\lim_{t\to\infty}z(t;x)=F(x)$ for each $x$, it follows that  
the system \fr{ODE-z-F} is a relaxation generating system.
\end{Example}

Roughly speaking a relaxation generating system is a seed for generating 
a relaxation process. This is refined as follows: 

\begin{Theorem}
\label{fact-induced-relaxation}
(induced relaxation process 1).
Consider the relaxation generating system \fr{XF-ODE} with
some initial condition. 
Then, introduce a family of graph immersions $\{(f_t,\xi_t)\}_{t\in I}$
into $\mbbR^{n+1}$ 
written in coordinates as 
$x\mapsto (x,F_{t}(x))\in\mbbR^{n+1}$, $(t\in I)$, that is,   
$f_{t}$ is associated with a function $F_{t}$. In addition,  
\begin{enumerate}
\item
  choose the function $F_{t}$ to be $F_{t}(x)=z(t;x)$ with $z(t;x)$
  being a solution to \fr{XF-ODE}, and 
\item
  let $y_{a}(t;x)$  be such that    
$$
y_{a}(t;x)
=\frac{\partial F_{t}(x)}{\partial x^{a}},\qquad
a=1,\ldots,n.
$$
\end{enumerate}
If the limit and differentiation are commute, 
then the image of the curve $t\mapsto (x,z(t;x),y(t;x),w(t;x))$
is a relaxation process in the sense of
Definition\,\ref{Definition-relaxation-process}:
$$
\lim_{t\to \infty}z(t;x)
=F(x),\qquad\mbox{and}\qquad
\lim_{t\to \infty}y_{a}(t)
=\frac{\partial F}{\partial x^{a}}(x),\quad
a=1,\ldots,n.
$$
\end{Theorem}

\begin{Proof}
  The statement on $\lim z(t;x)$ is nothing but
  the definition of relaxation generating system.  
Then a proof for $y_{a}(t)$ is given below. 
For all $x$, it follows from 
$$
\lim_{t\to\infty }F_{t}(x)
=F(x),
$$
and commutability of the limit and differentiation that
$$
\lim_{t\to\infty }y_{a}(t)
=\frac{\partial F}{\partial x^{a}}(x),\qquad a=1,\ldots,n.
$$
\qed
\end{Proof}

\begin{Remark}
  Note the following:
\begin{enumerate}
 \item  
  Given a relaxation generating system, the induced
relaxation process is obtained by applying    
Theorem\,\ref{fact-induced-relaxation}.
 \item
  For each $t\in\cI$ of this family of graph immersions,
  $(f_t,\xi_t)$ can be interpreted as a
  set of equilibrium states.  
\item
  For each $t\in\cI$ of this family of graph immersions,
  a flat connection is
  induced where the connection can be written as $\nabla_{t}$.
\item
  For each $t\in \cI$ of this family of graph immersions,  
  the geometric divergence is induced if $F_{t}(x)$ 
  is convex or concave with respect to $x$, 
  where this geometric
  divergence can be written as $D_{t}^{\G}$.  
\end{enumerate}
\end{Remark}

The following is a physical model explaining 
Theorem\,\ref{fact-induced-relaxation} in a physical language.  
The following example also shows how a physical system is translated into
its affine geometric description of thermodynamics proposed in this paper.

\begin{Example}
(kinetic Ising model without spin-coupling\,\cite{Goto2015JMP,Goto2020JMP}).
  Consider a nonequilibrium system consisting of only one spin
  $\sigma=\pm 1$ in contact with a heat bath of fixed temperature
  $T=1/(k_{\B}\beta)$ with $k_{\B}$ being the Boltzmann constant. The dynamics of the statistical average of this spin is governed 
  by an externally applied static magnetic field $\cH$, where 
  this $\cH$ yields a relaxation process towards a point of 
  the unique set of equilibrium states.   
  To describe this system,  introduce a time-independent
  variable $x:=\beta\mu_{\B} \cH\in\mbbR$. This $x$ is 
  a dimensionless variable for $\cH$ with $\mu_{\B}$ being the Bohr magneton.
  In the following after the equilibrium case is discussed, 
  the nonequilibrium case is discussed. 
  The equilibrium probability distribution function
  is assumed to be the canonical distribution,
  $$
  \mbbP_{\eq}(\sigma,x)
  =\exp(x\sigma)/Z(x),\quad
  \mbox{where}\quad
  Z(x)=\sum_{\sigma=\pm1}\e^{\,x\sigma}
  =2\cosh x.
  $$
  At equilibrium, one can introduce the negative of a dimensionless free-energy
  $F:\mbbR\ni x\mapsto F(x)\in\mbbR$ as 
  $$
  F(x)
  :=\ln Z(x)
  =\ln \cosh(x)+\ln 2,
  $$
  the magnetization at equilibrium is then written as 
\beqa
  \avg{\sigma}_{\eq}(x)
  &:=&\sum_{\sigma=\pm 1}\mbbP_{\eq}(\sigma,x)\sigma
  \non\\
  &=&\tanh(x)=\frac{\dr F}{\dr x}.
  \non
  \eeqa
  In the following a dynamical equation for this system is derived. 
  Let $I=\mbbR$. 
  Introduce $\mbbP_{t}:\{\pm1\}\times\mbbR \to \mbbR_{>0}$, 
  $(t\in \cI)$, that is  
  a probability distribution function of $\sigma$ at time $t$.  
   Then the expectation variable at $t$ is denoted by
  $\avg{\sigma}(t;x):=\sum_{\sigma=\pm1}\mbbP_{t}(\sigma;x)\,\sigma$,
   $(t\in\cI)$. This variable, $\avg{\sigma}(t;x)$,
   is identified with a nonequilibrium magnetization at $t$ for
   a fixed $x$.    
  Imposing the simple assumptions for $\mbbP_{t}$,
  \begin{itemize}
    \item the equation for $\mbbP_{t}$ is the Pauli master equation, and
    \item the detailed balance condition holds for the master equation, 
  \end{itemize}
  one derives the ODE: 
  $$
  \frac{\dr }{\dr t}\avg{\sigma}
  =\tanh(x)-\avg{\sigma}.
  $$
This system of the ODE is referred to as the
{\it kinetic Ising model without spin-coupling} in Ref.\,\cite{Goto2015JMP}.  
Solving this ODE explicitly, one verifies that 
  $$
  \lim_{t\to\infty}\avg{\sigma}(t;x)
  =\avg{\sigma}_{\eq}(x).
  $$
  To write this system in affine geometry, identify $y(t;x)=\avg{\sigma}(t;x)$,
  and 
  introduce the new variable $z(t;x)\in\mbbR$. The physical meaning of $z$ 
  is the value of a nonequilibrium extension of $F$, and $z(t;x)$ 
  is denoted by $F_t(x)$.   
  A relaxation generating system is introduced  by letting
  $w_{\,F}(t;x)=F(x)-z$ as in \fr{ODE-z-F},
  $$
  \frac{\dr z}{\dr t}
  =\ln\cosh x+\ln 2-z,
  $$
  whose solution \fr{ODE-z-F-solution} is
  written as $z=F_t(x)$ with   
  $$
  F_t(x)
  =(1-\e^{\,-t})(\ln\cosh(x)+\ln 2)+\e^{\,-t}z(0;x).
  $$
  The set $\{(f_{t},\xi_{t})\}_{t\in \cI}$  
  is a family of graph immersions, and the image of the curve
  $t\mapsto (x,y(t;x),z(t;x),w_{\,F}(t;x))$
  is a relaxation process.   
\end{Example}
 
\subsubsection{Nonequilibrium, (ii) case of two sets of equilibrium states }
\label{section-nonequilibrium-non-unique-equilibrium}

In the following, a nonequilibrium geometric theory is proposed for the case
that there are two sets of equilibrium states.
This is 
based on the geometric theories developed in
Sections\,\ref{section-geometric-equilibrium} and
\ref{section-nonequilibrium-unique-equilibrium}.
To avoid unnecessary confusion and to keep discussions in this paper simple,  
cases where more than three equilibrium states are not discussed in this paper.

The following is the definition of relaxation process
from a point of the set of metastable equilibrium states to
a point of the set of the most 
stable equilibrium states.  

\begin{Definition}
\label{Definition-relaxation-process-2}
(relaxation process with metastable state).
For
$\Omega_{0}\subset\Omega$,
let $x$ be coordinates of a point $q_{0}\in \Omega_{0}$,
$z\in\mbbR$, $\xi=\partial/\partial z$, and 
$y_{a}=\dr z(f_{*}(\partial/\partial x^{a}))$, ($a=1,\ldots,n$). 
In addition let $F_{\I}:\Omega_{0}\to\mbbR$ and $F_{\II}:\Omega_{0}\to \mbbR$
be functions where the condition $F_{\I}(q_{0})< F_{\II}(q_{0})$ holds for
any $q_{0}\in\Omega_{0}$. Consider the two graph immersions given below:  
\begin{itemize}
\item
  A graph immersion $(f_{\I},\xi)$ associated with
  $F_{\I}$ into  $\mbbR^{n+1}$.
  Suppose that the image of $(f_{\I},\xi)$ is identified with
  the most stable equilibrium state set. 
\item
  A graph immersion $(f_{\II},\xi)$ associated with
  $F_{\II}$ into  $\mbbR^{n+1}$.
  Suppose that the image of $(f_{\II},\xi)$ is identified with
  a metastable  equilibrium state set.  
\end{itemize}
In addition, let $\phi:\mbbR \ni t\mapsto p(t)\in T\mbbR^{n+1}$ be a curve, 
where the coordinates of $p(t)$ are denoted by 
$(x(t),z(t),y(t),w(t))$, ($t\in \mbbR$).
If a class of curves satisfies the conditions    

\begin{enumerate}
\setcounter{enumi}{-1}
\item
$$
  z(0)<F_{\II}(x),\quad\mbox{and}\quad 
  x(t)=x(0),\quad \forall\,t\in \mbbR,
$$
\item
$$
\lim_{t\to-\infty}z(t)
=F_{\II}(x),\quad\mbox{and}\quad
\lim_{t\to-\infty}y_{a}(t)
=\frac{\partial F_{\II}}{\partial x^{a}}(x),
\qquad a=1,\ldots,n,
$$
\item
$$
\lim_{t\to\infty}z(t)
=F_{\I}(x),\quad\mbox{and}\quad
\lim_{t\to\infty}y_{a}(t)
=\frac{\partial F_{\I}}{\partial x^{a}}(x), 
\qquad a=1,\ldots,n,
$$
\end{enumerate}
then the image of the 
curve is said to be a {\it relaxation process} from
a point of $(f_{\II},\xi)$ towards
a point of $(f_{\I},\xi)$. 
\end{Definition}

To state an affine geometric description of this class of relaxation processes,
two sets of equilibrium states are placed as follows.
Let $(f_{\I},\xi)$ and $(f_{\II},\xi)$ 
be graph immersions into $\mbbR^{n+1}$ 
written in coordinates as
$x\mapsto (x,F_{\I}(x))\in\mbbR^{n+1}$ and
$x\mapsto (x,F_{\II}(x))\in\mbbR^{n+1}$, respectively,
where $\xi=\partial/\partial z$.  
In addition, let $\Omega_{0}\subset\Omega$ be 
 a region on which the condition   $F_{\II}(x)>F_{\I}(x)$ is satisfied.   
First, recognize that $\Omega_{0}\times \mbbR\subset\mbbR^{n+1}$ is a trivial
fiber bundle, $\pi:\Omega_{0}\times\mbbR\to\Omega_{0}, \pi(q_{0},z)=q_{0}$.  
Hence $z\in\mbbR$ is a point of the fiber whose base point is $q_{0}$.  
Then, consider the vector field on the fiber $\mbbR$ over a point
$q_{0}\in\Omega_{0}\subset\Omega$ of the form written in coordinates 
\beq
X_{\II\to\I}
=w_{\,\II\to\I}(z;x)\frac{\partial}{\partial z}
\label{XF-II-I}
\eeq
with $w_{\,\II\to\I}(-;x)$ being a function of $z$ at a given $x$  
(see Fig.\,\ref{picture-relaxation-vector-fields} (Right)).  
Integral curves of \fr{XF-II-I} are obtained by solving the ODE 
\beq
\frac{\dr z}{\dr t}
=w_{\,\II\to\I}(z;x),\qquad t\in\mbbR.  
\label{XF-ODE-II-I}
\eeq
If a solution $z(t;x)$ whose initial condition is $z(0)<F_{\II}(x)$ 
satisfies 
$$
\mbox{1.}\ \lim_{t\to-\infty}z(t;x)
=F_{\II}(x),\qquad\mbox{and}\qquad
\mbox{2.}\ \lim_{t\to\infty}z(t;x)
=F_{\I}(x),
$$
then the system \fr{XF-ODE-II-I} is
referred to as a  
{\it relaxation generating system} with $f_{\I}$ and
$f_{\II}$,  
and the corresponding vector field $X_{\II\to\I}$  
{\it relaxation generating vector field} with $f_{\I}$ and
$f_{\II}$ in this paper. 

\begin{Example}
\label{Example-zeta=F-z-2}
Consider the dynamical system on $\mbbR$ associated with $X_{\II\to\I}$: 
\beq
\frac{\dr z}{\dr t}
=w_{\,\II\to\I}(z;x),\qquad
w_{\,\II\to\I}(z;x)
=-(z-F_{\I}(x))(z-F_{\II}(x))^{2},\quad t\in\mbbR, 
\qquad \mbox{for a fixed $x$}, 
\label{ODE-z-F-II-I}
\eeq
where the graph of $w_{\,\II\to\I}(z;x)$ is drawn
in Fig.\,\ref{figure-zeta-II-I}. 
In the case $z(0)<F_{\II}(x)$ in $\mbbR$ for a given $x$, 
it follows that 
\beq
\lim_{t\to-\infty}z(t;x)
=F_{\II}(x),\qquad\mbox{and}\qquad
\lim_{t\to\infty}z(t;x)
=F_{\I}(x).
\label{ODE-z-F-II-I-solution}
\eeq
\begin{figure}[htb]
\begin{center}
\includegraphics[scale=1.0]{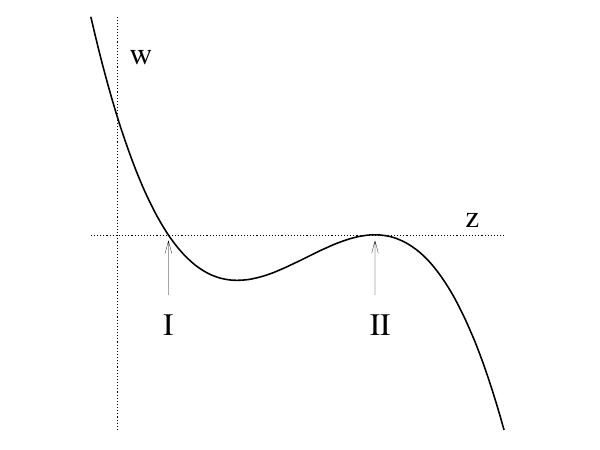}  
\end{center}
\caption{Graph of $w_{\,\II\to\I}(z;x)=-(z-F_{\I}(x))(z-F_{\II}(x))^2$.
  The symbols I and II in the graph indicate $F_{\I}$ and $F_{\II}$,
  respectively.}
\label{figure-zeta-II-I}
\end{figure}
Hence the system \fr{ODE-z-F-II-I} is a relaxation generating system.
\end{Example}
 There are various variants  of \fr{ODE-z-F-II-I}.  
 One of them is
$$
w_{\,\II\to\I}(z;x)
=(z-F_{\I}(x))(z-F_{\II}(x)).
$$

Roughly speaking a relaxation generating system is a seed for generating 
a relaxation process. This is refined as follows: 

\begin{Theorem}
\label{fact-induced-relaxation-2}
  (induced relaxation process 2).
Consider the relaxation generating system \fr{XF-ODE-II-I}. 
Then, introduce a family of graph immersions $\{(f_t,\xi_t)\}_{t\in \mbbR}$
into $\mbbR^{n+1}$ 
written in coordinates as 
$x\mapsto (x,F_{t}(x))\in\mbbR^{n+1}$, $(t\in \mbbR)$, that is,   
$f_{t}$ is associated with a function $F_{t}$.
\begin{enumerate}
  \item 
Choose $F_{t}$ to be 
$F_{t}(x)=z(t;x)$ with $z(t;x)$ being a solution to \fr{XF-ODE-II-I}, and 
\item
let $y_{a}(t;x)$ be   
$$
y_{a}(t;x)
=\frac{\partial F_{t}(x)}{\partial x^{a}},\qquad
a=1,\ldots,n.
$$
\end{enumerate}
If the limit and differential are commute, 
then the image of the 
 curve $t\mapsto (x,y(t;x),z(t;x),w_{\,\II\to\I}(t;x))$
is a relaxation process in the sense of
Definition\,\ref{Definition-relaxation-process-2}:
\begin{enumerate}
\item  
$$
\lim_{t\to -\infty}z(t;x)
=F_{\II}(x),\qquad\mbox{and}\qquad
\lim_{t\to -\infty}y_{a}(t;x)
=\frac{\partial F_{\II}}{\partial x^{a}}(t;x),\quad
a=1,\ldots,n.
$$
\item  
$$
\lim_{t\to \infty}z(t;x)
=F_{\I}(x),\qquad\mbox{and}\qquad
\lim_{t\to \infty}y_{a}(t)
=\frac{\partial F_{\I}}{\partial x^{a}}(t;x),\quad
a=1,\ldots,n.
$$
\end{enumerate}
\end{Theorem}

\begin{Proof}
  The statement on $\lim z(t;x)$ is nothing but the definition of relaxation
  generating  system.  
Then a proof for $y_{a}(t)$ is given below. 
For all $x$, it follows from 
$$
\lim_{t\to-\infty }F_t(x)
=F_{\II}(x),\qquad 
\lim_{t\to\infty }F_t(x)
=F_{\I}(x),
$$
and commutability of the limit and differentiation that
$$
\lim_{t\to-\infty }y_{a}(t)
=\frac{\partial F_{\II}}{\partial x^{a}}(x),\qquad
\lim_{t\to\infty }y_{a}(t)
=\frac{\partial F_{\I}}{\partial x^{a}}(x), 
 \qquad a=1,\ldots,n.
$$
\qed
\end{Proof}

\begin{Remark}
Given a relaxation generating system, the induced
relaxation process is obtained by applying 
Theorem\,\ref{fact-induced-relaxation-2}. Note the following:
\begin{itemize}
\item
  For each $t\in\mbbR$, 
  $(f_t,\xi_t)$ can be interpreted as
  a set of equilibrium states.
\item
   For each $t\in\mbbR$, 
  a flat connection is
  induced where the connection can be written as $\nabla_{t}$.

\item
  For each $t\in \cI$, 
  the geometric divergence is induced
  if $F_{t}$ is convex or concave with respect to $x$, where this geometric
   divergence can be written as $D_{t}^{\G}$.  
\end{itemize}
\end{Remark}

\section{Comparison with contact geometric thermodynamics}
\label{section-comparison}
In this section the developed theory in
Section\,\ref{section-geometric-description-thermodynamics}
of this paper is compared with
a representative existing theory.
As a representative thermodynamic theory
a contact geometric thermodynamics is summarized first in  
Section\,\ref{section-comparison-contact}. Second 
a lift of the vector field $X_{F}$ on $\mbbR$ is discussed in  
Section\,\ref{section-comparison-lift}. 
The reason why lifted vector fields are considered is that
the manifold on which $X_{F}$ is defined is different to the manifold 
on which contact vector fields are defined. 
To compare these vector fields on the equal footing,
one needs the same dimensional manifolds on which vector fields are defined.
One way to realize such is to lift the vector fields on a lower dimensional
manifold. For the same reason, a lift of $X_{\II\to\I}$ is discussed.  
Then, in Section\,\ref{section-comparison-compare}, 
the present affine geometric theory 
is compared with the existing  contact geometric theory.
For nonequilibrium systems,
vector fields are compared by introducing appropriate identifications
for different manifolds.

\subsection{Contact geometric thermodynamics} 
\label{section-comparison-contact}
One of developing geometric theories of thermodynamics employs
contact geometry, where contact geometry is known as an 
odd-dimensional counterpart of symplectic geometry.  
In the contact geometric thermodynamics, the manifold $T^{*}Q\times\mbbR$
is often considered as an ambient manifold\,\cite{Leon2020,Goto2022JMP}, where
$Q$ is an $n$-dimensional manifold. Let $x$ be a coordinate system of 
$Q$,  $y$ that of $T_{q}^{*}Q$, and $z$ that of $\mbbR$. In addition, 
equip the one-form
$\lambda=\dr z-\sum_{a=1}^{n}y_{a}\,\dr x^{a}$ on $T^{*}Q\times \mbbR$, where
the top-form $\lambda\wedge\dr\lambda\wedge\cdots\wedge\dr \lambda$ does not
vanish anywhere. An odd-dimensional manifold with $\ker\lambda$
  is called a {\it contact manifold} where $\ker\lambda$ denotes the kernel of $\lambda$. Then the 
pair $(T^{*}Q\times\mbbR,\ker\lambda)$ is a contact manifold.

A Legendrian submanifold of a contact manifold 
can be interpreted as a
thermodynamic phase space\cite{Mrugala2000},
where the equation of state  at equilibrium is described.
Here a {\it Legendrian submanifold} is 
an $n$-dimensional submanifold of a contact manifold 
satisfying the condition that the pull-back of 
 $\lambda$ vanishes. 
There are some useful projections\,\cite{Arnold-Givental}.
In terms of the coordinate system $(x,y,z)$,
the projection of a Legendrian submanifold 
onto the $(x,z)$-plane is called a {\it Legendre map},
its image is celled a {\it wave front}. 
In addition, the projection of a Legendrian submanifold
onto the $(x,y)$-plane is called a {\it Lagrange map}. 

A diffeomorphism that preserves the contact structure $\ker\lambda$
is called a {\it contact transform}, and  
its vector field is called a {\it contact vector field}. 
This vector field is sometimes employed as a tool for expressing 
nonequilibrium time-evolution of thermodynamic systems
\cite{Goto2015JMP,Goto2022JMP,Entov2021,GLP2022} 
(see Ref.\, \cite{Jurkowski2000} for another thermodynamic  
interpretation of a contact vector field).   
A way to provide a
contact vector field is to provide a function, called a  
{\it contact Hamiltonian},
and the derived contact vector field is called a 
{\it contact Hamiltonian vector field}.
A contact Hamiltonian vector field $Y_{h}$ associated with contact
Hamiltonian $h$ 
is determined by
$$
\ii_{Y_{h}}\lambda
=h,\qquad\mbox{and}\qquad
\ii_{Y_{h}}\dr\lambda
=-\dr h+(Rh)\lambda,
$$
where $R$ is called the {\it Reeb vector field} defined such that
$$
\ii_{R}\lambda
=1,\qquad\mbox{and}\qquad
\ii_{R}\dr \lambda=0.
$$
In the case where the contact form is expressed as above,
the contact Hamiltonian vector field $Y_{h}$ on $T^{*}Q\times\mbbR$
is written in coordinates as 
$$
Y_{h}
=\sum_{a=1}^{n}\left(\dot{x}^{a}\frac{\partial }{\partial x^{a}}
+\dot{y}_{a}\frac{\partial }{\partial y_{a}}\right) 
+\dot{z}\frac{\partial }{\partial z}.
$$
By identifying $\dot{\ }=\dr/\dr t$, one derives 
$$
\frac{\dr}{\dr t}x^{a}
=-\frac{\partial h}{\partial y_{a}},\qquad
\frac{\dr}{\dr t}y_{a}
=y_{a}\frac{\partial h}{\partial z}+\frac{\partial h}{\partial x^{a}},
\qquad 
\frac{\dr }{\dr t}z
=h-\sum_{b=1}^{n}y_{b}\frac{\partial h}{\partial y_{b}},\quad
 a=1,\ldots,n.
$$
When $h$ does not depend on $y$, $h=h(x,z)$, one immediately has that 
\beq
\frac{\dr}{\dr t}x^{a}
=0,\qquad
\frac{\dr}{\dr t}y_{a}
=y_{a}\frac{\partial h}{\partial z}+\frac{\partial h}{\partial x^{a}},
\qquad 
\frac{\dr }{\dr t}z
=h,\quad a=1,\ldots,n.
\label{contact-Hamiltonian-x-z}
\eeq
The following two contact Hamiltonian systems will be
  focused in  Section\,\ref{section-comparison-compare}.  
\begin{itemize}
  \item
Choose $h$ to be $h_{F}$, where 
$$
h_{F}(x,z)
=F(x)-z,
$$
with $F$ being some function. 
Substituting this $h_F$ into \fr{contact-Hamiltonian-x-z}, one has 
\beq
\frac{\dr x^{a}}{\dr t}
=0,\qquad
\frac{\dr y_{a}}{\dr t}
=-y_{a}+\frac{\partial F}{\partial x^{a}},
\qquad 
\frac{\dr z}{\dr t}
=h_{F},\quad a=1,\ldots,n.
\label{contact-Hamiltonian-F-z}
\eeq
The contact Hamiltonian system \fr{contact-Hamiltonian-F-z}
  has been studied in Refs.\,\cite{Goto2015JMP,Goto2020Scripta}, so that
  a class of relaxation processes is described on a contact manifold.

\item
  Choose $h$ to be $h_{\II\to\I}$, where 
$$
h_{\II\to\I}(x,z)
=-(z-F_{I}(x))(z-F_{\II}(x))^{2},
$$
with $F_{\I}$ and $F_{\II}$ being functions of $x$.
Substituting this $h_{\II\to\I}$ into \fr{contact-Hamiltonian-x-z}, one has 
\beq
\frac{\dr x^{a}}{\dr t}
=0,\quad
\frac{\dr y_{a}}{\dr t}
=-(z-F_{\II})^{2}\left(y_{a}-\frac{\partial F}{\partial x^{a}}\right)
-2(z-F_{\I})(z-F_{\II})\left(y_{a}-\frac{\partial F_{\II}}{\partial x^{a}}\right),
\quad 
\frac{\dr z}{\dr t}
=h_{\II\to\I},
\label{contact-Hamiltonian-F-z-2}
\eeq
where $a=1,\ldots,n$.
The contact Hamiltonian system \fr{contact-Hamiltonian-F-z-2}
has been studied in Ref.\,\cite{Goto2022JMP}, so that
a class of relaxation processes with sets of multiple equilibrium states
is described on a contact manifold.
In Ref.\,\cite{Goto2022JMP},
the dimension of the contact manifold is three, and
suffix $a$ in \fr{contact-Hamiltonian-F-z-2} has been omitted. 
\end{itemize}

\subsection{Lift of relaxation generating vector field}
\label{section-comparison-lift}
In this subsection 
vector fields on a higher dimensional manifold are defined
based on the developed theories in Section\,\ref{section-nonequilibrium}, 
so that the vector fields on the higher dimensional manifold
will be compared with the existing vector fields
defined on a contact manifold. 
In particular, a vector field lifted from $X_{F}$ in \fr{XF} and
a vector field lifted from $X_{\II\to\I}$ in \fr{XF-II-I} 
are discussed here,
where the lifted vector fields are denoted by $\wt{X_{F}}$ and
$\wt{X_{\II\to\I}}$, respectively. 
Recall that 
the vector fields $X_{F}$ and $\wt{X_{F}}$ generate relaxation processes for 
the unique equilibrium state systems,
and that vector fields $X_{\II\to\I}$ and $\wt{X_{\II\to\I}}$ generate
relaxation processes for non-unique equilibrium state systems.

Before discussing lifted vector fields for the both cases, 
how to discuss stability of
vector fields on a manifold $\cM$ of dimension $n$
is summarized here.  
Let $\mu$ be a volume-element of $\cM$, and $X$ a vector field on 
$\cM$.
Then the {\it phase space compressibility} of $X$ with respect to $\mu$, 
denoted $\ddiv_{\mu}X\in\GamLam{\cM}{0}$, is
defined such that\,\cite{Ezra2002} 
$$
\cL_{X}\mu
=(\ddiv_{\mu} X)\mu, 
$$
where $\cL_{X}$ is the Lie derivative along $X$. 
This function is sometimes considered when stability of
vector fields on manifolds is discussed\,\cite{Goto2016JMP}. 
If the system associated with $X$  has the property   
that $(\ddiv_{\mu}X)(p)<0$ at $p\in\cM$,  
then this dynamical system is said to be {\it contracting} at
$p\in \cM$ with
respect to $\mu$ in this paper.
A coordinate expression of $\ddiv_{\mu}X$ is obtained as follows. 
Let $\xi=(\xi^{1},\ldots,\xi^{n})$ be coordinates, and $\mu$ be  
such that
$$
\mu=\dr\xi^{1}\wedge\cdots\wedge\dr\xi^{n}.
$$
Note that the coordinates $\xi$ are 
nothing to do with any transversal vector field of
an affine immersion. 
Consider the vector field written in coordinates $\xi$ of the form 
$$
X=\sum_{a=1}^{n}\dot{\xi}^{a}(\xi)\frac{\partial}{\partial \xi^{a}},
$$
where $\dot{\xi}=(\dot{\xi}^{1},\ldots,\dot{\xi}^{n})$
is a set of functions of $\xi$.  
Then the Lie derivative of $\mu$ along $X$ is calculated to be
$$
\cL_{X}\mu
=\underbrace{(\dr\dot{\xi}^{1}\wedge\dr\xi^{2}\wedge\cdots\wedge\dr\xi^{n})
  +\cdots+
(\dr\xi^{1}\wedge\cdots\wedge\dr\xi^{n-1}\wedge\dr\dot{\xi}^{n})}_{n\,\mathrm{terms}}
=\sum_{a=1}^{n}\frac{\partial \dot{\xi}^{a}}{\partial \xi^{a}}\mu,
$$
from which one can write $\ddiv_{\mu}X$ as a function of $\xi$, 
$$
(\ddiv_{\mu}X)(\xi)
=\sum_{a=1}^{n}\frac{\partial \dot{\xi}^{a}}{\partial \xi^{a}}.
$$
To see the role of this function more clearly,
consider the one-dimensional case,
$X=\dot{z}\partial/\partial z$ on $\cM=\mbbR$ with $\mu=\dr z$ and
$\dot{z}=-z$. Integral curves of $X$ are obtained by solving the ODE  
$\dot{z}=-z$, where $\dot{z}=\dr z/\dr t$. 
It follows from 
the explicit solution for $z$ that  
the absolute value $|z(t)|$ is decreasing (contracting) 
as time develops from $t=0$ to any $t_{*}>0$.  
Meanwhile, for this system
the phase space compressibility as a function of $z$ is 
obtained by the calculation $\partial(-z)/\partial z=-1$ as
$(\ddiv_{\dr z}X)(z)=-1$ on $\mbbR$, and one concludes that
this system is contracting.  
This example justifies the terminology ``contracting.''   

Below, the case of the unique equilibrium set and 
the case of non-unique equilibrium state sets 
are discussed separately. 

\subsubsection{Unique set of equilibrium states}
In Theorem\,\ref{fact-induced-relaxation},
the variables $y$ and $z$ evolve in time.  
Meanwhile the corresponding vector field $X_{F}$ and the dynamical system are
written only in terms of $z$ for a fixed $q$ and $x$. 
Hence it is natural to consider 
a dynamical system involving $y$ and $z$, and is natural to consider 
its corresponding vector 
field denoted by $\wt{X_{F}}$.
This vector field is studied here. 
Since there are various classes
of ODEs, one focuses on systems where 
\begin{enumerate}
\item
a fixed point $(y_{*\ a},z_{*})$
is  $(\partial F/\partial x^{a},F(x))$, $(a=1,\ldots,n)$
with $x$ being a set of coordinate values of a point
$q\in\Omega$, and 
\item 
the corresponding vector field is a lift of $X_{F}$, in the sense that 
$\pi_{*}\wt{X_{F}}=X_{F}$: 
$$
\xymatrix{
  T\mbbR
  & T(T_{q}\Omega \times\mbbR )
    \ar[l]_(0.66){\pi_{*}}\\
  \mbbR\ar[u]^{X_F}
  &  T_{q}\Omega\times  \mbbR\ar[u]_{\wt{X_F}}
  \ar[l]_(0.65){\pi}
},\qquad
\xymatrix{
  (z,\dot{z})&(y,\dot{y},z,\dot{z})\ar@{|->}[l]_(0.55){\pi_{*}}\\
  z\ar@{|->}[u]&(y,z)\ar@{|->}[u]\ar@{|->}[l]_(0.5){\pi}
}.
$$
Note that the manifold $T\Omega\times\mbbR$ has been discussed 
in \fr{projection-Tf} as a redundant manifold for describing equilibrium
states.
\end{enumerate}
This $\wt{X_{F}}$ is written of the form: 
\beq
\wt{X}_{F}
=\sum_{a=1}^{n}u_{\,F\,a}(z,y;x)\frac{\partial}{\partial y_{a}}+
w_{\,F}(z;x)\frac{\partial}{\partial z},
\label{tilde-XF}
\eeq
with some functions $u=(u_{\,F\,1},\ldots,u_{\,F\,n})$. 
The reason why item 1 is needed is as follows.
If $(\partial F/\partial x^{a},F)$  is a fixed point and attractive 
in some sense,
then integral curves of $\wt{X_{F}}$
are relaxation processes, which we seek. 
Here  relaxation process
in $T(T_{q}\Omega\times \mbbR)$ is defined as follows.

\begin{Definition}
\label{Definition-relaxation-process-lifted}
(relaxation process in the lifted space 1).   
Let $\phi:I\ni t\mapsto p(t)\in T(T_{q}\Omega\times\mbbR)$ 
be a curve, where the coordinates of $p(t)$ are denoted by 
$(y(t),z(t),\dot{y}(t),\dot{z}(t))$, ($t\in I\subset \mbbR$).
For each $x$ kept fixed in $t$, if a curve satisfies the conditions    
\beq
\lim_{t\to\infty}y_{a}(t)
=\frac{\partial F}{\partial x^{a}},\quad\mbox{and}\quad
\lim_{t\to\infty}z(t)
=F(x),\qquad a=1,\ldots,n,
\label{conditions-relaxation-process-lifted}
\eeq
then the image of the 
curve is said to be a {\it relaxation process} 
towards a point of the set of equilibrium states $(f,\xi)$.
\end{Definition}

To compare $X_{F}$ with $\wt{X_{F}}$, a property of $X_{F}$ is 
studied first. 
In particular, 
a stability of the dynamical system of the ODE \fr{XF-ODE}
associated with $X_{F}$ 
on $\mbbR$, $\dot{z}=w_{\,F}(z;x)$,
is discussed here. 
Let  $\ddiv_{\dr z}\,X_{F}\in\GamLam{\mbbR}{0}$ 
be the phase space compressibility with respect to $\dr z$
on $\mbbR$, that is, 
$$
\cL_{X_{F}}\dr z
=(\ddiv_{\dr z}\,X_{F})\dr z.  
$$
Simple calculations yield 
$$
(\ddiv_{\dr z}\,X_{F})(z;x)
=\frac{\partial w_{\,F}}{\partial z}(z;x).
$$
  
To study properties of $\wt{X_{F}}$, 
similar to the case of $X_{F}$,  
one defines the function
$\ddiv_{\mu}{\wt{X_{F}}}\in \GamLam{(T_{q}\Omega\times\mbbR)}{0}$
  such that
$$
\cL_{\wt{X_{F}}}\mu
=(\ddiv_{\mu}\,\wt{X_{F}})\mu,\quad\mbox{where}\quad
\mu=\dr y_{1}\wedge\cdots\wedge \dr y_{n}\wedge \dr z.
$$

The following Lemma is about the system of the ODEs
associated with $X_{F}$ 
and that with $\wt{X_{F}}$.
When $(y_{*\,a},z_{*})=(\partial F/\partial x^{a},F(x))$ is a fixed point
for the system of the ODEs associated with $\wt{X_{F}}$  
and
conditions for Theorem\,\ref{fact-induced-relaxation} are satisfied, 
this fixed point is an attractor for the dynamical system.  
\begin{Lemma}
\label{fact-fixed-point-z-y-generating-system}
Consider a dynamical system for $(y,z)$ associated with $\wt{X_{F}}$, 
\beq
\frac{\dr z}{\dr t}
=w_{\,F}(z;x),\qquad  
\frac{\dr y_{a}}{\dr t}
=u_{\,F\,a}(z,y;x),\qquad
a=1,\ldots,n,
\label{ODE-z-y}
\eeq
where $u_{\,F\,a}(-,-;x)$ is a function of $z$ and $y$ for a given $x$.
Choose 
\beq
u_{\,F\,a}(z,y;x)
=y_{a}\frac{\partial w_{\,F}}{\partial z}
+\frac{\partial w_{\,F}}{\partial x^{a}},\qquad a=1,\ldots,n.
\label{ODE-z-y-Upsilon}
\eeq
Let $z_{*}$ be a point that satisfies $w_{\,F}(z_{*};x)=0$, i.e., 
the point $z_{*}$ is a fixed point for the ODE $\dot{z}=w_{\,F}$ in
\fr{ODE-z-y}. Then the following hold. 
\begin{enumerate}
\item
The contracting property is preserving under the lift in the sense that  
\beq
(\ddiv_{\dr z} X_{F})(z;x)
< 0\quad \mbox{on}\ \cZ
\qquad\Rightarrow\qquad 
(\ddiv_{\mu}\,\wt{X_{F}})(z,y;x)
< 0,\quad \mbox{on}\ T_{q}\Omega\times \cZ
\label{ddiv-X_F}
\eeq  
with some $\cZ\subset\mbbR$.
\item
  In a subset  $\cZ\subset \mbbR$ containing $z_{*}$, 
  if $w_{\,F}(z;x)\geq 0$, $\partial w_{\,F}/\partial z\leq 0$, and 
$\partial w_{\,F}/\partial z|_{z_{*}}= 0$,
  then $\lim_{t\to\infty}(y,z)=(y_{*},z_{*})$.
\item
 If
\beq
\frac{\partial w_{\,F}}{\partial z}(z_{*};x)
=-1,
\label{ODE-z-y-fixed-point-condition-z}
\eeq
then $y_{*}$ of a fixed point $(y_{*},z_{*};x)$ is given by
\beq
y_{*\,a}
 =\frac{\partial w_{\,F}}{\partial x^{a}}(z_{*};x),\qquad
 a=1,\ldots,n.
\label{ODE-z-y-fixed-point-y}
\eeq
\end{enumerate}
\end{Lemma}
\begin{Proof}
(Proof for 1.)
  The explicit form of $\ddiv_{\mu}\,\wt{X_{F}}$ is obtained as
$$
(\ddiv_{\mu}\,\wt{X_{F}})(z,y;x)
=(n+1)\frac{\partial w_{\,F}}{\partial z}(z;x).
$$
From this and the condition
$(\ddiv_{\dr z}\,X_{F})(z;x)=\partial w_{\,F}/\partial z< 0$,
item 1 holds. 

\noindent
(Proof for 2.) This follows from the Theorem of Lyapunov\,\cite{smale}.

\noindent
(Proof for 3.)
Substituting  \fr{ODE-z-y-fixed-point-condition-z}
  into \fr{ODE-z-y-Upsilon}, one has 
  $$
  u_{\,F\,a}(z_{*},y_{*};x)
  =y_{*\,a}\left.\frac{\partial w_{\,F}}{\partial z}\right|_{*}
  +\left.\frac{\partial w_{\,F}}{\partial x^{a}}\right|_{*}
=-y_{*\,a}  +\left.\frac{\partial w_{\,F}}{\partial x^{a}}\right|_{*},
  \qquad a=1,\ldots,n.
  $$
  To find an explicit form of a fixed point for  
  $\dot{y}_{a}=u_{\,F\,a}$,
  letting  
$\dot{y}_{*\,a}=u_{\,F\,a}(z_{*},y_{*}; x)=0$ for each $x$, one has 
\fr{ODE-z-y-fixed-point-y}.
\qed
\end{Proof}

In Lemma\,\ref{fact-fixed-point-z-y-generating-system}, note the following. 
\begin{itemize}
\item
  In Example\,\ref{Example-zeta=F-z},
  the case $w_{\,F}(z;x)=F(x)-z$ has been considered. 
  In this example it follows 
  that $(\ddiv_{\dr z} X_{F})(z;x)=-1$.
   Hence the condition for item 1 and that for 3 are satisfied.
  In addition, 
  the subset $\cZ$ for item 2 is found as $\cZ=\{z|F(x)-z\geq 0\}$.
  These calculations are summarized as Tab.\ref{table-unique-stable}.
  \begin{table}[htb]
  \caption{Behavior of the system in Example\,\ref{Example-zeta=F-z} }
  \begin{center}
    \begin{tabular}{|c||c|c|c|}
      \hline
      $z$   &  & $F(x)$  & \\
      \hline\hline
      $w_{\,F}(z;x)$ & $+$ & $0$ & $-$\\
      \hline
      $(\ddiv_{\dr z}X_{F})(z;x)$ & $-1$ & $-1$&$-1$\\
      \hline
    \end{tabular}
  \end{center}
  \label{table-unique-stable}
 \end{table}
  
\item
  The system consisting of \fr{ODE-z-y}, \fr{ODE-z-y-Upsilon}, and
  $\dot{x}^{a}=0$ $(a=1,\ldots,n)$, 
  is formally the same as
  \fr{contact-Hamiltonian-F-z},
    where \fr{contact-Hamiltonian-F-z} is  
  the contact Hamiltonian system with the contact
  Hamiltonian $h(x,z)=F(x)-z$. 
  Hence, 
  it is expected that there is a relation between the lifted  
  relaxation generating systems and contact Hamiltonian systems. 
  In particular, the function $w_{\,F}(z;x)$ is expected to play a role of
  $h(x,z)$.  
  This role will be discussed in
  Section\,\ref{section-comparison-compare}.
\end{itemize}

The following Proposition
shows how to describe a relaxation process in terms of
$\wt{X_{F}}$.  

\begin{Proposition}
\label{fact-lifted-system-yield-relaxation}
  Consider the dynamical system
  \fr{ODE-z-y} with \fr{ODE-z-y-Upsilon}.
In the case where 
$$
w_{\,F}(z;x)
=F(x)-z,
$$
with an initial condition $F(x)>z(0)$,
integral curves of $\wt{X_{F}}$ connect points on
$T_{q}\Omega\times\mbbR$ 
and the fixed point $(\partial F/\partial x^{a},F(x))$.
\end{Proposition}
\begin{Proof}
  This follows from item 2 of
  Lemma\,\ref{fact-fixed-point-z-y-generating-system}.
\end{Proof}

Lemmata\,\ref{fact-fixed-point-z-y-generating-system}
and \ref{fact-lifted-system-yield-relaxation} are associated with 
Theorem\,\ref{fact-induced-relaxation} that is
the case of a unique set of equilibrium states.  
Meanwhile the following are associated with
Theorem\,\ref{fact-induced-relaxation-2}. 

\subsubsection{Two sets of equilibrium states}
In Theorem\,\ref{fact-induced-relaxation-2},
the variables $y$ and $z$ evolve in time. 
Meanwhile the corresponding vector field $X_{\II\to\I}$
and the dynamical system are
written only in terms of $z$ for a fixed $q_{0}$ and $x$. 
Hence it is natural to consider 
a dynamical system involving $y$ and $z$, and its corresponding vector
field denoted by $\wt{X_{\II\to\I}}$. 
This vector field is studied here. 
Since there are various classes
of ODEs, one focuses on systems where 
\begin{enumerate}
\item
  a fixed point $(y_{*\, a},z_{*})$ is 
  $(\partial F_{\I}/\partial x^{a},F_{\I}(x))$, 
  $(a=1,\ldots,n)$, with $x$ being the coordinate system for $\Omega_{0}$, 
\item 
  the corresponding vector field is a lift of $X_{\II\to\I}$,
  in the sense that $\pi_{*}\wt{X_{\II\to\I}}=X_{\II\to\I}$: 
$$
\xymatrix{
  T\mbbR
  & T(T_{q_{0}}\Omega_{0} \times\mbbR )
    \ar[l]_(0.66){\pi_{*}}\\
  \mbbR\ar[u]^{X_{\II\to\I}}
  &  T_{q_{0}}\Omega_{0}\times  \mbbR\ar[u]_{\wt{X_{\II\to\I}}}
  \ar[l]_(0.65){\pi}
},\qquad
\xymatrix{
  (z,\dot{z})&(y,\dot{y},z,\dot{z})\ar@{|->}[l]_(0.55){\pi_{*}}\\
  z\ar@{|->}[u]&(y,z)\ar@{|->}[u]\ar@{|->}[l]_(0.5){\pi}
}.
$$
Note that the manifold $T\Omega\times\mbbR$ has been discussed 
in \fr{projection-Tf} as a redundant manifold for describing equilibrium
states.
\end{enumerate}
This $\wt{X_{\II\to\I}}$ is written of the form: 
\beq
\wt{X_{\II\to\I}}
=\sum_{a=1}^{n}u_{\,\II\to\I\,a}(z,y;x)\frac{\partial}{\partial y_{a}}+
w_{\,\II\to\I}(z;x)\frac{\partial}{\partial z},
\label{tilde-X-II-I}
\eeq
with some functions
$u_{\,\II\to\I}=(u_{\,\II\to\I\,1},\ldots,u_{\,\II\to\I\,n})$. 
The reason why item 1 is needed is as follows.
If $(\partial F_{\I}/\partial x^{a},F_{\I})$  is a fixed point and attractive 
in some sense,
then integral curves of $\wt{X_{\II\to\I}}$
are relaxation processes, which we seek. 
Here relaxation process
in $T(T_{q_{0}}\Omega_{0}\times \mbbR)$ is defined as follows.

\begin{Definition}
\label{Definition-relaxation-process-lifted-2}
(relaxation process in the lifted space 2). 
Let $\phi:\mbbR \ni t\mapsto p(t)\in T(T_{q_{0}}\Omega_{0}\times\mbbR)$
be a curve, where the coordinates of $p(t)$ are denoted by 
$(y(t),z(t),\dot{y}(t),\dot{z}(t))$, ($t\in\mbbR$).
For each $x$ kept fixed in $t$, if a curve satisfies the conditions    
\begin{enumerate}
\item
\beq
\lim_{t\to-\infty}y_{a}(t)
=\frac{\partial F_{\II}}{\partial x^{a}},\quad\mbox{and}\quad
\lim_{t\to-\infty}z(t)
=F_{\II}(x),\qquad
a=1,\ldots,n,
\label{conditions-relaxation-process-lifted-2-}
\eeq
\item
\beq
\lim_{t\to\infty}y_{a}(t)
=\frac{\partial F_{\I}}{\partial x^{a}},\quad\mbox{and}\quad
\lim_{t\to\infty}z(t)
=F_{\I}(x),\qquad
a=1,\ldots,n,
\label{conditions-relaxation-process-lifted-2}
\eeq
\end{enumerate}
then the image of the 
curve is said to be a {\it relaxation process} 
towards a point of the set of equilibrium
states $(f,\xi)$.
\end{Definition}

To compare $X_{\II\to\I}$ with $\wt{X_{\II\to\I}}$, a property of
$X_{\II\to\I}$ is studied first. 
In particular, 
a stability of the dynamical system of the ODE \fr{XF-ODE-II-I}
associated with $X_{\II\to\I}$ on $\mbbR$, $\dot{z}=w_{\,\II\to\I}(z;x)$,
is discussed here. 
Similar to the case of $\ddiv_{\dr z}\,X_{F}\in\GamLam{\mbbR}{0}$,  
let $\ddiv_{\dr z}\,X_{\II\to\I}\in\GamLam{\mbbR}{0}$ 
be the phase space compressibility with respect to $\dr z$ on $\mbbR$,
that is, 
$$
\cL_{X_{\II\to\I}}\dr z
=(\ddiv_{\dr z}\,X_{\II\to\I})\dr z.  
$$
Simple calculations yield 
$$
(\ddiv_{\dr z}\,X_{\II\to\I})(z,y;x) 
=\frac{\partial w_{\,\II\to\I}}{\partial z}(z;x).
$$
If $\partial w_{\,\II\to\I}/\partial z<0$ in some $\cZ\subset\mbbR$, 
then  the system is contracting in $\cZ$.  
  
To study properties of $\wt{X_{\II\to\I}}$, 
similar to the case of $X_{\II\to\I}$, 
one defines the function
$\ddiv_{\mu}{\wt{X_{\II\to\I}}}
\in \GamLam{(T_{q_{0}}\Omega_{0}\times\mbbR)}{0}$
such that
$$
\cL_{\wt{X_{\II\to\I}}}\mu
=(\ddiv_{\mu}\,\wt{X_{\II\to\I}})\mu,\quad\mbox{where}\quad
\mu=\dr y_{1}\wedge\cdots\wedge \dr y_{n}\wedge \dr z.
$$

The following Lemma is about the system of the ODEs associated
with $X_{\II\to\I}$ and that with $\wt{X_{\II\to\I}}$. 
\begin{Lemma}
\label{fact-fixed-point-z-y-generating-system-2}
Consider a dynamical system for $(y,z)$ associated with $\wt{X_{\II\to\I}}$, 
\beq
\frac{\dr z}{\dr t}
=w_{\,\II\to\I}(z;x),\qquad  
\frac{\dr y_{a}}{\dr t}
=u_{\,\II\to\I\,a}(z,y;x),\qquad
a=1,\ldots,n,
\label{ODE-z-y-2}
\eeq
where $u_{\,\II\to\I\,a}(-,-;x)$ is a function of $z$ and $y$
for a given $x$.
Choose 
\beq
u_{\,\II\to\I\,a}(z,y;x)
=y_{a}\frac{\partial w_{\,\II\to\I}}{\partial z}
+\frac{\partial w_{\,\II\to\I}}{\partial x^{a}},\qquad a=1,\ldots,n.
\label{ODE-z-y-Upsilon-2}
\eeq
Then the contracting property is preserving under the lift in the sense that  
\beq
(\ddiv_{\dr z} X_{\II\to\I})(z;x)
< 0\quad \mbox{on}\ \cZ
\qquad\Rightarrow\qquad 
(\ddiv_{\mu}\,\wt{X_{\II\to\I}})(z,y;x)
< 0,\quad \mbox{on}\ T_{q_{0}}\Omega_{0}\times \cZ,
\label{ddiv-X_F-2}
\eeq  
with some $\cZ\subset\mbbR$.
\end{Lemma}
\begin{Proof}
  A way to prove this is analogous to the proof of item 1 of 
   Lemma\,\ref{fact-fixed-point-z-y-generating-system}.
\end{Proof}

In Lemma\,\ref{fact-fixed-point-z-y-generating-system-2}, note the following. 
\begin{itemize}
\item
  In Example\,\ref{Example-zeta=F-z-2},
  the case $w_{\,\II\to\I}(z;x)=-(z-F_{\I}(x))(z-F_{\II}(x))^2$
  has been considered. 
  In this example it follows 
  that $(\ddiv_{\dr z} X_{\II\to\I})(z;x)< 0$ in some 
  $\cZ\subset\mbbR$. 
  Hence there is some non-empty $\cZ$ such that the condition for
  Lemma\,\ref{fact-fixed-point-z-y-generating-system-2}  
  is satisfied. 
  More precisely, one has 
  $$
  (\ddiv_{\dr z}X_{\II\to\I})(z;x)
  =\left\{
  \begin{array}{cl}
    \mbox{positive}& z_{0}(x)<z<F_{\II}(x)\\
         0&z=z_{0}(x),\quad z=F_{\II}(x)\\
    \mbox{negative}& \mbox{otherwise}
  \end{array}
  \right.\qquad 
  z_{0}(x):=\frac{2F_{\I}(x)+F_{\II}(x)}{3}. 
  $$
  Note that $F_{\I}(x)<z_{0}<F_{\II}(x)$ due to 
  $F_{\I}(x)-z_{0}(x)=(F_{\I}(x)-F_{\II}(x))/3<0$ and
  $F_{\II}(x)-z_{0}(x)=(2/3)(F_{\II}(x)-F_{\I}(x))>0$.
  These calculations
  are summarized in Tab.\,\ref{table-metastable}.
From this summary, it follows that
  $(\ddiv_{\dr z} X_{\II\to\I})(z;x)<0$ and
$(\ddiv_{\mu} \wt{X_{\II\to\I}})(z,y;x)<0$
around $z=F_{\I}(x)$. 
  \begin{table}[htb]
  \caption{Behavior of the system in Example\,\ref{Example-zeta=F-z-2} }
  \begin{center}
    \begin{tabular}{|c||c|c|c|c|c|c|c|}
      \hline
      $z$          &  & $F_{\I}(x)$  &
      & $z_{0}(x)$ &  & $F_{\II}(x)$ &  \\
      \hline\hline
      $w_{\,\II\to\I}(z;x)$ & $+$ & $0$ & $-$
      & $-$&$-$&$0$&$-$ \\
      \hline
      $(\ddiv_{\dr z}X_{\II\to\I})(z;x)$ & $-$ & $-$&$-$
      &$0$ &$+$ &$0$ &$-$\\
      \hline
    \end{tabular}
  \end{center}
  \label{table-metastable}
 \end{table}

  Moreover, in this Example the point $(y_{*\,a},z_{*})$ in the system
  \fr{ODE-z-y-2} with \fr{ODE-z-y-Upsilon-2},
   $$
  (y_{*\,a}(x),z_{*}(x))
  =\left(\frac{\partial F_{\I}}{\partial x^{a}}(x),F_{\I}(x)\right),\quad
  a=1,\ldots,n,
  $$
  is verified to be a fixed point. 

\item
  The system consisting of \fr{ODE-z-y-2}, \fr{ODE-z-y-Upsilon-2}, and
  $\dot{x}^{a}=0$ $(a=1,\ldots,n)$, 
  is formally the same as \fr{contact-Hamiltonian-F-z-2}, where
\fr{contact-Hamiltonian-F-z-2}  is
  the contact Hamiltonian system with the contact
  Hamiltonian $h(x,z)=-(z-F_{I}(x))(z-F_{\II}(x))^{2}$.  
  Hence, 
  it is expected that there is some relation between the lifted  
  relaxation generating systems and contact Hamiltonian systems. 
  In particular, the function $w_{\,F}(z;x)$ is expected to play a role of
  $h(x,z)$.  
  This role will be discussed in
  Section\,\ref{section-comparison-compare}.
\end{itemize}

The following Proposition
shows how to describe a relaxation process in terms of
$\wt{X_{\II\to\I}}$.  

\begin{Proposition}
\label{fact-lifted-system-yield-relaxation-2}
  Consider the dynamical system
  \fr{ODE-z-y-2} with \fr{ODE-z-y-Upsilon-2}.
In the case of Example\,\ref{Example-zeta=F-z-2}, 
$$
w_{\,\II\to\I}(z;x)
=-(z-F_{\I}(x))(z-F_{\II}(x))^{2},
$$
with an initial condition $z(0)<F_{\II}(x)$,
integral curves of $\wt{X_{\II\to\I}}$ connect points on
$T_{q_{0}}\Omega_{0}\times\mbbR$
and the fixed point $(\partial F_{\I}/\partial x^{a},F_{\I}(x))$.
\end{Proposition}

\begin{Proof}
  The strategy for proving this Lemma is to find Lyapunov functions, 
  ans this proof is 
  similar to the proof of Theorem 3.1 in Ref.\,\cite{Goto2022JMP}.
  
Let $V_{\I}(-;x)$ and 
$V_{\II}(-;x)$ be the functions for a given $x$, 
\beqa
V_{\I}(z;x)
&=&\frac{1}{2}(z-F_{\I}(x))^2,\qquad
\mbox{on}\quad \cZ_{\I}(x)=\{z|z<F_{\II}(x)\},\quad
\mbox{and}
\non\\
V_{\II}(z;x)
&=&z-F_{\II}(x),\qquad \mbox{on}\quad
\cZ_{\II}(x)
=\{z| F_{\II}(x)\leq z\}.
\non
\eeqa
It follows that $V_{\I}(z;x)\geq 0$ on $\cZ_{\I}(x)$, and that 
$$
\frac{\dr V_{\I}}{\dr t}
=\dot{z}(z-F_{\I}(x))
=-(z-F_{\I}(x))^{2}(z-F_{\II}(x))^{2}
\leq 0,\quad\mbox{on}\ \cZ_{\I}(x). 
$$
The equality holds when $z=F_{\I}(x)$. 
Hence $V_{\I}(-;x)$ is a Lyapunov function on $\cZ_{\I}(x)$.
Next,
it follows that $V_{\II}(z;x)\geq 0$ on $\cZ_{\II}(x)$, and that 
$$
\frac{\dr V_{\II}}{\dr t}
=\dot{z}
=-(z-F_{\I})(z-F_{\II}(x))^{2}
\leq 0,\quad\mbox{on}\ \cZ_{\II}(x). 
$$
The equality holds when $z=F_{\II}(x)$. 
Hence $V_{\II}(-;x)$ is a Lyapunov function on $\cZ_{\II}(x)$. 

Applying the Theorem of Lyapunov, one completes the proof.
\qed
\end{Proof}

\subsection{Comparisons}
\label{section-comparison-compare} 

In the following, the present affine geometric thermodynamics is 
compared with a contact geometric thermodynamics 
for equilibrium and nonequilibrium systems. 
Note that contact geometric thermodynamics is an developing branch of mathematical physics,
and these identifications may differ among theories. 

\subsubsection{Equilibrium}
Table\,\ref{table-comparison} shows identifications of notions
used in equilibrium thermodynamics  in {the languages of} contact geometry and
affine geometric thermodynamics.  

\begin{table}[htbp]
  \caption{Comparisons between affine and contact geometric thermodynamics}
\centering
  \begin{tabular}{|c||c|c|}
    \hline
    Thermodynamics& Contact geometric theory & Affine geometric theory \\
    \hline\hline
    Primal variables $x$ & Coordinates of $Q$
    & Coordinates of $\Omega$\\
    \hline
    Conjugate variables $y$ & Coordinates of $T_{q}^{*}Q$
    &  Some of coordinates of $T_{f(q)}\mbbR^{n+1}$ \\
\hline    
Complete function $F$ & A function on $Q$ & $F$ for a graph immersion \\
 \hline
 Fundamental relation& Contact form& Conormal map \\
 \hline
State equation & Legendrian submfd & Graph immersion \\
\hline
Response function & Hessian of $F$ & Affine fundamental form $h$ \\
 \hline
 Graph $(x,F)$ & Image of Legendre map & Image of graph immersion \\
 \hline
 Graph $(x,y)$ & Image of Lagrange map & (Unnamed projection)\\
\hline
  \end{tabular}
  \label{table-comparison}
\end{table}

\subsubsection{Nonequilibrium}
Notice that the system
\fr{contact-Hamiltonian-x-z} is formally the same as
\fr{ODE-z-y} with \fr{ODE-z-y-Upsilon}, and is also formally the same as
\fr{ODE-z-y-2} with \fr{ODE-z-y-Upsilon-2}.
In addition,
since $x$ is a coordinate of a point $q$ or $q_{0}$,   
one recognizes that
the equations for $y$ and $z$ in \fr{contact-Hamiltonian-x-z} are coordinate
expressions for 
the restricted vector field $Y_{h}|_{q}\in\GT{(T_{q}^{*}Q\times\mbbR)}$
at $q\in Q$.

One then immediately arrives at the following.
\begin{Theorem}
\label{fact-comparison-vector-fields}
  (relation between vector fields in affine and contact geometries).
A class of contact Hamiltonian vector fields 
$Y_{h}|_{q}\in \GT{(T_{q}^{*}Q\times \mbbR})$ 
is formally the same as the lifted relaxation generating vector field
$\wt{X_{F}}\in \GT{(T_{q}\Omega\times \mbbR)}$ in \fr{tilde-XF} 
with \fr{ODE-z-y} and \fr{ODE-z-y-Upsilon}, and is formally the  
same as $\wt{X_{\II\to\I}}\in \GT{(T_{q}^{*}Q\times \mbbR})$
 in \fr{tilde-X-II-I} with \fr{ODE-z-y-2} and \fr{ODE-z-y-Upsilon-2}.
How to identify $Y_{h}|_{q}$ with $\wt{X_{F}}$ and $\wt{X_{\II\to\I}}$ is 
to put $Q=\Omega$ or $Q=\Omega_{0}$ and identify
$T^{*}Q\cong T\Omega$ or $T^{*}Q\cong T\Omega_{0}$. 
\end{Theorem}

From Theorem\,\ref{fact-comparison-vector-fields}, it follows that 
the present affine geometric formalism is consistent with a 
 contact geometric formalism\,\cite{Goto2015JMP,Goto2022JMP,Entov2021}.
 In addition, this Theorem indicates 
 how a contact vector field is constructed
 from a given relaxation generating vector field on $\mbbR$. 
 Conversely,  a relaxation generating vector field 
 on $\mbbR$ is easily obtained 
 from a contact Hamiltonian vector field on a contact manifold by defining an
 appropriate projection.  
\section{Concluding remarks}
\label{section-Conclusions}

This paper offers an affine geometric
formulation of thermodynamic systems. This formulation
covers both equilibrium and simple nonequilibrium systems. 
The main claims of this paper are as follows:
\begin{itemize}
\item
  A set of equilibrium states is identified with
  the image of a graph immersion into $\mbbR^{n+1}$
  (see Interpretation\,\ref{interpretation-basic}).  
\item
  Several affine geometric objects can be introduced in thermodynamics
  (see Proposition\,\ref{fact-equilibrium-thermodynamics-correspondence-1}).  
\item
  A class of relaxation processes in thermodynamic systems are described
  (see Theorems\,\ref{fact-induced-relaxation} and
\ref{fact-induced-relaxation-2}).
\item
  The present geometric formulation of 
  relaxation processes are consistent with the existing theory 
  (see Theorem\,\ref{fact-comparison-vector-fields}). 
\end{itemize}
The significance of this study includes, as indicated the diagram below,
$$
\xymatrix@C=50pt@R=40pt{
  &  \mbox{\fbox{Affine Geom}}
  \ar@{-}[dr]|{\mbox{\cite{Kurose1994,Matsuzoe2010,Uohashi2000}}}
  \ar@{.}[dl]|{\mbox{Present study}}
  &\\
  \mbox{\fbox{Thermodynamics}}&
  &\mbox{\fbox{Information Geom}}
  \ar@/_13pt/@{-}[ll]|{\mbox{\cite{Wada2015,Nakamura2019,Sagawa2022}}}  \\
  &\mbox{\fbox{Contact Geom}}
  \ar@{-}[ur]|{\mbox{\cite{Mori2018,Nakajima2021,Goto2015JMP}}}
  \ar@{-}[ul]|{\mbox{\cite{Harmann1973,Mrugala2000,Gromov2011,Bravetti2019,Grmela2014,Lopez2021,Haslach1997}}}
  \ar@{-}[ul]\ar@{.}[uu]|{\small\mbox{Present study}}
  &\mbox{\fbox{Symplectic Geom}}
  \ar@{-}[u]|{\mbox{\cite{Nielsen1997,Noda2011,favretti2020entropy}}}
  \ar@{-}[l]|{\mbox{\cite{Arnold,Libermann1987,Silva2008}}}
 }
$$
\begin{itemize}
\item
  shedding light on a potential link between
  affine geometry and thermodynamics, so that
  several methodologies in affine geometry will be introduced to the study of
  thermodynamics.  
\item
  showing how to compare vector fields on
  an extended affine immersion with those
  on a contact manifold, which provides relations  
  between the study of
  affine geometry and that of contact geometry. 
\end{itemize}

There remain unsolved problems that have not been addressed in this paper.
They include
\begin{itemize}
\item
  showing applications of the present approach to
  various thermodynamic systems
  and electric circuits\,\cite{Maschke2006,Goto2016JMP},  
\item 
  showing applications of the affine immersion theory of codimension two to
  thermodynamic systems \cite{Matsuzoe2010},

\item
  exploring relations between affine geometry and statistical
  mechanics\,\cite{Ezra2002,Mrugala1990PRA,Bravetti2019}, 
  
\item 
  exploring relations among affine geometry, symplectic and Liouville
  geometries,
  since symplectic and Liouville geometries are
  employed to describe  
  thermodynamics
  \cite{Schaft2018entropy,favretti2005entropy}.
\end{itemize}
By addressing these, it is expected that a relevant and sophisticated
geometric methodology will be established for dealing with 
various thermodynamic systems and related systems. 
In addition, related mathematics are expected to be developed.

 \section*{Acknowledgment}
 The author was partially supported by
 JSPS (KAKENHI) grant number JP19K03635, and thanks
Minoru Koga at Nagoya University  for giving
suggestions and fruitful discussions on this study.  
The author also thanks Hiroshi Matsuzoe at Nagoya
Institute of Technology for giving suggestions related to this study. 

\subsection*{Conflict of interest}
The author has no conflicts to disclose.

 \subsection*{Data Availability Statement}
Data sharing is not applicable to this article as no new data were created or analyzed in this study.



\begin{thebibliography}{99}
\bibitem{Callen} 
  H.B. Callen, 
  {\it Thermodynamics and an Introduction to Thermostatistics, 2nd Edition},
  Wiley, 1985

\bibitem{Frenkel} 
  T. Frenkel, 
  {\it The Geometry of Physics, 3rd edition},
  Cambridge University Press,  2011

\bibitem{Nakahara} 
  M. Nakahara,
  {\it Geometry, topology and physics, 2nd edition},  
  CRC press, 2003

\bibitem{Harmann1973} 
R. Harmann, {\it Geometry, physic and systems}, 
Dekker, 1973

\bibitem{Mrugala2000} 
  R. Mrugala,  
  On contact and metric structures on thermodynamic spaces,
  {\it Suken kokyuroku}, 
  {\bf 1142}, 167–181, 2000    

\bibitem{Baldiotti2016} 
  M.C. Baldiotti, R. Fresneda, C. Molina,
  A Hamiltonian approach to Thermodynamics
  {\it Ann. Phys.} {\bf 373}, 245--256, 2016 

\bibitem{Schaft2018entropy}
  A. Schaft, and B. Maschke, 
  Geometry of thermodynamic processes,
  {\it Entropy }, {\bf 20}, 925, 2018
  
\bibitem{Yoshimura2017} 
  F. Gay-Balmaz and H. Yoshimura,
A Lagrangian variational formulation for nonequilibrium
thermodynamics. Part I: Discrete systems,  
{\it J. Geo. Phys.}, {\bf 111}, 194--212, 2017
  
\bibitem{Arnold}
  V.I. Arnold, {\it Mathematical Methods of Classical Mechanics}, 
  Springer, 1978  

\bibitem{Libermann1987}
  P. Libermann, and C-M. Marle,
  {\it Symplectic Geometry and Analytical Mechanics},
  Springer, 1987

\bibitem{Silva2008}
  A. da Silva, {\it  Lectures on Symplectic Geometry},
  Springer, 2008

\bibitem{McInerney2013}
  A. McInerney, 
  {\it First Steps in Differential Geometry},   
  Springer, 2013

\bibitem{Arnold-Givental} 
  V.I. Arnold and A.B. Givental, 
  Symplectic geometry, {\it Dynamical Systems IV, 
    Symplectic Geometry and its Applications},
  edited by V.I. Arnold, 
  and S. Novikov, Encyclopedia of Mathematical Sciences 4, Springer, 1990 

\bibitem{Weinhold1976} 
F. Weinhold,
Metric geometry of equilibrium thermodynamics,
{\it J. Chem. Phys.}, {\bf 63}, 2479–2483, 1975
  
\bibitem{Ruppeiner1995} 
  G. Ruppeiner,
Metric geometry of equilibrium thermodynamics
{\it Rev. Mod. Phys.}, {\bf 67}, 605–659, 1995
  
\bibitem{NomizuSasaki1994}
   K. Nomizu and T. Sasaki, 
   {\it Affine Differential Geometry: Geometry of Affine Immersions} 
   (Cambridge Tracts in Mathematics, Series Number {\bf 111}),
   Cambridge University Press, 1994 

\bibitem{Matsuzoe2010}
   H. Matsuzoe,
  Statistical manifolds and affine differential geometry,
  {\it Advanced Studies in Pure Mathematics}, {\bf 57},
  Mathematical Society of Japan, 
  2010

\bibitem{AmariNagaoka2000} 
  S.I. Amari and H. Nagaoka, Methods of Information Geometry,
  Translations of Mathematical Monographs, {\bf 191}
  (American Mathematical Society, Providence, 2000)

\bibitem{Kurose1994}
  T. Kurose, 
  On the divergences of 1-conformally flat statistical manifolds,
  {\it T\^ohoku  Math. J.}, {\bf 46}, 427–433, 1994
  
\bibitem{Wada2015}
  T. Wada, H. Matsuzoe, A.M. Scarforne,
  Dualistic Hessian Structures Among the Thermodynamic Potentials
  in the $\kappa$-Thermostatistics,
  {\it Entropy}, {\bf 17}, 7213-7229, 2015

\bibitem{Nakamura2019}
  T. Nakamura, H.H. Hasegawa, and D.J. Driebe,  
  Reconsideration of the generalized second law based on information geometry,
{\it J.Phys. Commun.}, {\bf 3}, 015015, 2019  

\bibitem{Sagawa2022} 
   T. Sagawa, 
  {\it Entropy, Divergence, and Majorization in Classical and Quantum Thermodynamics}, Springer, 2022

\bibitem{Shima2007}
  H. Shima,
  {\it The geometry of Hessian Structures}, World Scientific, 2007

\bibitem{Sasa2000}
  S. Sasa,
  {\it Introduction to Theremodynamics}, 2000 (In Japanese), 
  Kyoritu shuppan, 2000

  
\bibitem{Kubo1991}
  R. Kubo, M. Toda, and N. Hashitsume, 
  {\it Statistical Physics II}, Springer, 1991

\bibitem{Zubarev1996}
D. Zubarev, V. Morozov, and G. Ropke, 
{\it Statistical Mechanics of Nonequilibrium Processes, Basic Concepts, Kinetic Theory}, Weily, 1996

\bibitem{Mrugala1978} 
  R. Mrugala,  
  Geometrical formulation of equilibrium phenomenological thermodynamics,
  {\it Rep. Math. Phys.}, 
  {\bf 14}, 419–427, 1978   

\bibitem{Brody1995}
  D. Brody and N. River,
  Geometrical aspects of statistical mechanics,
  {\it Phys. Rev. E}, {\bf 51}, 1006-1011, 1995
  


\bibitem{Goto2015JMP}
  S. Goto,
  Legendre submanifolds in contact manifolds as  attractors and
  geometric nonequilibrium thermodynamics,
  {\it J. Math. Phys.}, {\bf 56}, 073301, 2015

\bibitem{Goto2020JMP}
  S. Goto, and H. Hino,
  Diffusion equations from master equations - A discrete geometric approach,
  {\it J. Math. Phys.}, {\bf 61}, 113301, 2020
   
\bibitem{Leon2020} 
  A.A. Simoes, M. de Leon, M.L. Valcazar, D.M. de Diego, 
  Contact geometry for simple thermodynamical systems with friction,
  {\it Proc. Roy. Soc. A}, {\bf 476}, 20200244, 2020   

\bibitem{Goto2022JMP}
  S. Goto,
Nonequilibrium thermodynamic process with hysteresis and metastable states—A contact Hamiltonian with unstable and stable segments of a Legendre submanifold,
  {\it J. Math. Phys.}, {\bf 36}, 053302, 2022

\bibitem{Entov2021}
  M. Entov, and L. Polterovich,
  Contact topology and non-equilibrium thermodynamics,
  {\tt arXiv:2101.037701} 
  

\bibitem{GLP2022}
  S. Goto, S. Lerer, and L. Polterovich,
  Contact geometric approach to Glauber dynamics near cusp and its limitation,
  J. Phys. A: Math.Theor., {\bf 56}, 125001, 2023 ({\tt arXiv:2210.00703}) 
  
\bibitem{Jurkowski2000}  
  J. Jurkowski,
  Canonical deformations of surfaces of equilibrium states in thermodynamic
  phase space,
  {\it Phys. Rev. E}, {\bf62}, 1790-1798, 2000

\bibitem{Goto2020Scripta} 
  S. Goto and H. Hino,
  Information and contact geometric description of expectation variables
  exactly derived from master equations,
  {\it Phys. Scr.}, {\bf 95} 015207, 2020
  
\bibitem{Ezra2002}
  G.S. Ezra,
  Geometric approach to response theory in non-Hamiltonian systems,
  {\it J. Math. Chem.}, {\bf 32}, 339--360, 2002

\bibitem{Goto2016JMP}
  S. Goto,  
  Contact geometric descriptions of vector fields on dually flat spaces and
  their applications in electric circuit models and nonequilibrium statistical
  mechanics
  {\it J. Math. Phys.}, {\bf 57}, 102702, 2016
  
\bibitem{smale}
  M.W. Hirsch and S. Smale,
  {\it Differential Equations, Dynamical Systems, and Linear Algebra},
  Academic Press, 1974

\bibitem{Uohashi2000}
   K. Uohashi, A. Ohara, T. Fujii,
   Foliations and divergences of flat statistical manifolds,
   {\it Hiroshima Math. J.}, {\bf 30}, 403-414, 2000

\bibitem{Grmela2014}
  M. Grmela,
  Contact geometry of mesoscopic thermodynamics and dynamics,
  {\it Entropy}, {\bf 16}, 1652–1686, 2014

\bibitem{Gromov2011}
  D. Gromov, and P.E. Cairns, 
  Interconnection of thermodynamic control systems,
 {\it IFAC Proc.} {\bf 44}, 6091--6097, 2011  
  
\bibitem{Mori2018} 
  A. Mori,  
  Information geometry in a global setting,
  {\it Hiroshima Math. J.}, {\bf 48}, 291--305, 2018
  
\bibitem{Nakajima2021}
  N. Nakajima, and T. Ohmoto, 
  The dually flat structure for singular models,
  {\it Info. Geo.}, {\bf 4}, 31, 2021 

\bibitem{favretti2020entropy} 
M. Favretti, 
Lagrangian Submanifolds of Symplectic Structures Induced
by Divergence Functions,  
{\it Entropy}, {\bf 22},  983 [13pages], 2020

\bibitem{Bravetti2019}
  A. Bravetti,
  Contact geometry and thermodynamics,   
  {\it Int. J. Geo. Methods, Mod. Phys.},{\bf 16}, 1940003 [51pages], 2019 


\bibitem{Lopez2021}
  C.S. Lopez-Monsalvo, F. Nettel, V. Pineda-Reyes,
  and L.F. Escamilla-Herrera,
  Contact polarizations and associated metrics in geometric thermodynamics,
  {\it J. Phys. A: Math. Theor.}, {\bf 54}, 105202, 2021 

\bibitem{Nielsen1997} 
  O.E. Barndorff-Nielsen and P.E. Jupp,
  Statistics, yokes and symplectic geometry,
  {\it Ann. Fac. Sci. Toulouse Math.}, {\bf 6}, 389–427, 1997
  
\bibitem{Noda2011}
  A. Noda, Symplectic structures on statistical manifolds, 
  {\it J. Aust. Math. Soc.}, {\bf 90}, 371–384, 2011

\bibitem{Haslach1997}
  H.W. Haslach,  
  Geometric structure of the non-equilibrium thermodynamics of homogeneous
  systems, 
  {\it Rep. Math. Phys.}, {\bf 39}, 147--162, 1997  

\bibitem{Maschke2006}
D. Eberard, B.M. Maschke and A.J. van der Schaft,
Energy-conserving formulation of RLC-circuits with linear resistors,
in Proceedings of the 17th International Symposium on Mathematical
Theory of Networks and Systems, Kyoto, Japan, 24–28 July, 2006  

\bibitem{Mrugala1990PRA} 
  R. Mrugala, J.D. Nulton, J.C. Schon, and  P. Salamon, 
Statistical approach to the geometric structure of thermodynamics,   
  {\it Phys. Rev. A}, {\bf 41}, 3156--3160, 1990   

\bibitem{favretti2005entropy}
M. Favretti,   
Lagrangian submanifolds generated by the Maximum Entropy principle,
{\it Entropy}, {\bf 7},  1--14, 2005
  






  
  

  

  





  




  



  





  


\end{thebibliography}
\end{document}